\documentclass[11pt, a4paper]{article}
\usepackage[a4paper, top=2.5cm, bottom=2.5cm, left=2cm, right=2cm]{geometry}
\usepackage{tikz}
\usepackage{caption}
\usepackage{subcaption}
\usepackage[square, numbers]{natbib}
\usepackage{abstract}
\usepackage{authblk}
\usepackage{url} 

\newcommand{\fourtops}{$t\bar{t}t\bar{t}$ }
\newcommand{\ttH}{$t\bar{t}H$ }
\newcommand{\ttbar}{$t\bar{t}$ }

\title{Weight-Based Representation Learning for Parameter Inference in Monte Carlo Simulations}

\author[1]{V. Wachirapusitanand}
\author[1]{N. Srimanobhas}

\affil[1]{\normalsize{Center of Excellence in High Energy Physics, Department of Physics, Faculty of Science, Chulalongkorn University, Bangkok 10330, Thailand}}

\begin{document}

\maketitle

\begin{abstract}
We present a Machine Learning-based approach for parameter inference in physics models that exploits event-level weights provided by simulators. Individual observations may have weights assigned by a simulation framework that describe the change in probability with respect to the model parameters. As these assigned weights encode the sensitivity of the parameter, they can serve as a weak supervision signal for learning parameter-informative representations. In this work, our inference models are trained using simulator-provided weights to learn representations and their relations to the parameter-sensitive structures in the high-dimensional observations. The resulting representations are then discretised into summary statistics and the model parameter value is inferred using a likelihood-based inference procedure. We illustrate this approach by using simulated four-top-quark production to infer the top quark Yukawa coupling (the parameter of interest). 
\end{abstract}

\section{Introduction}
A physics model is, in essence, a mathematical model that attempts to explain phenomena in physics. Since it is a mathematical model, it may contain a number of parameters that require tuning and measurements. For any physics model, a measurement of one parameter in the model may involve a large set of other parameters. While the physics models are already explicit by design, they may be intractable for measurements of certain parameters. Worse, numerical integration required for such measurements requires an enormous phase space of other latent parameters in the model, rendering the integration impossible.

Traditionally, this problem is tackled by relying on one or a few features residing in the data from a simulator, which generates high-dimensional observations $x \sim p(x|\theta) $ based on a set of continuous parameters $\theta$ used in the model. From these features, the parameter of interest may be inferred directly without learning, since the parameter can alter these features present in simulated observations $x_o$ and their statistical distributions. The inference can be done simply via the use of Bayesian inference to obtain the probable ranges of the parameter of interest $p(\theta|x_o)$~\cite{deistler}.

This approach, however, poses limitations to the parameter inference. First of all, the inference for the parameter of interest inside such a complex physics model may require an enormous set of simulated observables to properly represent a vast phase space inside the physics model. Simulating observables at this scale requires a large amount of computing power, storage, and time. The second limitation is that it is impossible to simulate observations over a continuous range of the parameter of interest. One needs to choose discrete points of the parameter value within the range and perform simulations over the chosen points.

On the other hand, in physics models, the probability distribution $p(x|\theta)$ in which the simulated observation is generated may be sensitive to the parameter of interest, resulting in some regions of the observation phase space being more pronounced than others. This effect can be realized by simulating a large batch of observations with different values of the parameter itself. However, to observe any difference in the observation induced by the parameters, the simulation may run into limitations, as we have discussed earlier.

To alleviate this, some simulators have a capability of providing weights for each observation $x$, which are dependent on the continuous parameters set in the model.
The weights assigned to each observation can alter the distribution of features in the same way as in weighted histograms. For physics models, observable weights can signify the sensitivity of the parameter of interest, as larger weight values for one particular observation can result in that observation becoming more pronounced in that particular observation phase space, and, by extension, in the distribution of features. The weights calculated by simulators are only available from the simulation and do not alter the features obtainable from one particular observation.

By calculating weights for observations based on the set of different values of parameters, we can replace multiple simulations from different values of parameters with one set of simulations, thereby reducing the computational resources used. Weight calculations for observations can also be used to improve the accuracy of the simulation itself by simulating the observations at lower accuracy and applying weights to those observations to better reflect the simulation at higher accuracy~\cite{Mattelaer_2016}.

The parameter inference can also be done by utilizing a machine learning model that derives informative representations via learning from the high-level features present in the observations $x$ from the simulator. The representation could be as simple as an output from a neural network, but this representation can contain essential information sufficient for the inference of the parameter. The inference can still be done using Bayesian inference, but by using the representation derived and simplified from the larger set of input features, the inference for the parameter of interest can be improved. The downside of this approach, however, is that although it can utilize a large set of input features derived from the observations, it disregards other valuable information not present in the observation itself, especially the information obtainable only at the simulation level.

A new approach for parameter inference, as proposed by Brehmer et al.~\cite{Brehmer_2018}, suggests that, in addition to relying on a small set of features available in data samples from simulations, we can exploit additional information from the simulation itself. The exploited information can also be used to train neural networks, and, in turn, be used to construct an unbinned likelihood ratio, and finally, the constrained regions of the model parameter of interest. The work shown in Ref.~\cite{Brehmer_2018} also presented several advanced approaches that can be used to construct the likelihood ratio. 

The simplest approach, as presented in Ref.~\cite{Brehmer_2018}, involves a histogram of features, where a template histogram is constructed from one or a few features from the dataset. With this approach, the contents of each histogram bin may change as the model parameter of interest varies, provided that the features are well-chosen. This approach is further adopted in a physics analysis~\cite{TOP-21-001}, where instead of using a histogram constructed from input features available in the dataset, a set of neural networks is trained to distinguish two sets of simulated data samples. The two datasets used by the work in Ref.~\cite{TOP-21-001} are simulated separately based on the different values of the model parameters of interest, and the training labels are simply determined based on the parameter configuration in the model. 

The existing approach, and its application in Ref.~\cite{TOP-21-001}, although being novel in terms of parameter inference, still relies on specifying the exact value of the parameter of interest during simulations. This results in multiple simulation sets required to cover the continuous range of the parameter of interest. If more simulated observations are required for more point values of the parameter of interest, the computational resources required for inference can scale up. An inference for multiple parameters utilizing this approach would also require an exponentially larger number of simulations to cover enough parameter variations, therefore requiring even more computational resources.

Here, we present an extension of the histogram-based approach where, instead of simulating two completely separated samples with different values of the model parameters of interest, we simulate one or more sets of observations and exploit the weights calculated from the simulation. As the weights are only available from a simulation, this approach effectively utilizes the latent information provided by the simulation in addition to the features extracted from the simulated observation. The exploit over observation weights should allow us to train better Machine Learning models to discriminate between observations with different behavior induced by different parameter values.

As a proof of concept, we will apply this new approach by constructing an inference model for the inference of the top Yukawa coupling, a parameter used in the Standard Model (SM), based on the simulation of four top quark production ($t\bar{t}t\bar{t}$). We will explain, in subsequent sections, the learning models used in the exploit as well as the likelihood-based inference methods, benchmarking the effectiveness of this new approach compared to the traditional approach of parameter inference via a surrogate quantity.

%\section{Section title}
%Sample text inserted for demonstration. Organize the main text of your article using section headings, and include any equations, figures, tables, lists etc using your preferred \LaTeX\ packages and commands. Example code for a figure and a table is given below, but you do not have to use this format. For general guidance on using \LaTeX , including information on figures, tables, equations and references, please refer to documents such as the \LaTeX\ WikiBook: \href{https://en.wikibooks.org/wiki/LaTeX}{https://en.wikibooks.org/wiki/LaTeX}.

%Note that clarity of presentation is the most important consideration when preparing your article for submission. It is not necessary to format your article in the style used for published articles in the journal.

%\subsection{Subsection title}
%Sample text inserted for demonstration, including links to figure \ref{fig1} and table \ref{tab1}.

%\subsubsection{Subsubsection heading}
%Sample text inserted for demonstration.

\section{Problem setup and data representation}
According to SM, the top quark Yukawa coupling ($y_t$) dictates the interaction between a Higgs boson and a top quark by altering the probability of any particle production involving this interaction to occur inside particle collisions, such as in particle colliders~\cite{Peskin:1995ev}. This is reflected by the quantity called the cross-section of the particle production. For any particle production, SM can provide a theoretical prediction of its cross section, which mainly depends on the centre-of-mass energy of the collision (such as 13~TeV for the LHC), the type of particles used in the collision (such as proton-proton collisions in the LHC), and other theoretical parameters, including the top quark Yukawa coupling. 

As explained in the previous section, this work will focus on the \fourtops production, which is a particle production resulting in two pairs of top-antitop quarks. This production has been discovered by both ATLAS~\cite{ATLAS-fourtops-observation} and CMS~\cite{TOP-22-013} Collaborations, and results from both collaborations show that the predictions provided by the SM still hold. Furthermore, prior to the two discovery results, CMS Collaboration~\cite{TOP-18-003} has already inferred the probable upper limit of $y_t$ by comparing the probable upper limit of \fourtops cross section, measured with collision data, against the SM prediction with respect to $y_t$. Meanwhile, ATLAS Collaboration~\cite{ATLAS-fourtops-observation} performed the inference of $y_t$ value by adjusting the \fourtops event yields along with event yields from other background productions with respect to $y_t$. In either case, the inference for probable regions of $y_t$ as a parameter of interest has not been the main priority and is performed based on the surrogate distribution of \fourtops event yields designed for the discovery of the production alone.

\subsection{Simulator-based learning with event reweighting}
Simulators in the High-Energy physics domain can simulate observations, or collision events in this case, based on the interactions defined in SM and parameter values set during the simulation. In a typical High-Energy physics analysis, the events collected from actual particle colliders and detectors are analyzed and compared to the simulated events from the simulators. Simulated events are crucial to the analyses since, unlike simulation, latent information for actual events is unobtainable, as it is impossible to determine the particle interactions that occur inside the collisions. On the other hand, we can simulate events explicitly based on particle productions of interest. A typical analysis would require simulated events for particle productions of interest (signal) and simulated events for particle productions that may provide the same signature as in the signal events.

\begin{figure*}
    \centering
    \includegraphics[width=\linewidth]{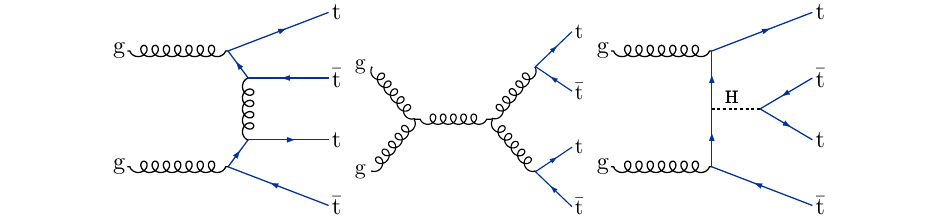}
    \caption{Feynman diagrams for \fourtops production at leading order, detailing the most probable events to occur from this particle production. Notice the rightmost diagram containing interactions between top quarks and Higgs bosons. Figure derived from Ref.~\cite{TOP-17-009}.}
    \label{fig:fourtops-diagrams}
\end{figure*}

One characteristic of \fourtops production is that, during the simulation, there is a possibility that a Higgs boson may occur during the production, as shown in the Feynman diagrams in Figure~\ref{fig:fourtops-diagrams}. If a simulated event based on this production contains the interactions between top quarks and Higgs bosons, the probability of the event being drawn from the simulator should differ if $y_t$ is altered. Furthermore, the addition of the Higgs boson in the simulated event may alter the kinematic features of the final state particles in the event. This means that the alteration of $y_t$, the parameter of interest, may affect both the weights of the event and the kinematic features of the event itself. 

Based on both changes imposed by $y_t$ to the event, we can build a Machine Learning model to exploit the connection between the kinematic features and the changes of the event weights induced by $y_t$. The model will mainly learn from the simulated events from the simulator, as they are the only set of events with the information necessary for learning.

\subsection{Dataset generation}
Collision events are generated using a chain of simulators that mimic the physical processes that occur inside particle detectors, starting from particle collisions inside proton beams, particle decays, radiation, and recombinations, interactions of resulting particles with detector hardware, digitization, and interpretation from (simulated) digitized detector signals to meaningful physics objects. 

Events generated based on \fourtops production are accompanied by a set of per-event weights, calculated at certain values of the parameter of interest $y_t$. This allows us to calculate the event yields with respect to $y_t$ and interpolate such yields into a continuous $y_t$ range for its inference.

To create a realistic inference setting, simulations of background processes are also required in addition to simulations for signal processes. The dominant background processes simulated in this work are:
\begin{itemize}
    %\item \ttbar with dileptonic decay and generator-level filter requiring $H_T > 500$ GeV and 7 particle jets or more.
    \item Top-antitop ($t\bar{t}$) production decaying into two leptons with a generator-level filter requiring a large number of particle jets and high total energy. This requirement is used to ensure that simulations for this background process can be represented without statistical issues where too few events pass the preselection criteria.
    %\item \ttH with $H \rightarrow b\bar{b}$ decay. This particular decay pathway ensures that events with two or more $b$-tagged jets can pass the preselection criteria.
    \item \ttH production with Higgs boson decaying into two $b$ quarks. This particular requirement ensures that events from this simulation can pass the preselection criteria with a sufficient number of events.
\end{itemize}

To make the learned model useful for the inference of $y_t$ using actual observations from particle detectors, the learning process must be based only on observables available at the reconstructed levels, while the observation weights provided by the simulator will be used as a target feature for the learning model.

\subsection{Feature representation and preselection} \label{subsec:feature-representation}
Each event generated from the simulation contains the information of final state particles, or particles resulting from the entire simulated process. The information from these final state particles is used to derive kinematic parameters and event-shape features, and these derived features are used as a representation for each event.

To make the inference for $y_t$ based on these datasets as realistic as in a typical data analysis in the High-Energy physics domain, several criteria are imposed on the simulated collision events to be included in the dataset. For a collision event to be included in the dataset, it is required to have exactly two leptons with different electric charges, 4 particle jets or more, two of them (or more) tagged as jets originating from $b$ quarks (so-called $b$-tagged jets)~\cite{DeepJet}. This collision event selection criterion is similar to another analysis aimed at the discovery of \fourtops production~\cite{TOP-21-005}, albeit with slightly more stringent requirements.

After the selection process, each event will have the input features extracted from the kinematic parameters of reconstructed final state particles, as well as event-shape features calculated from the entire set of particles present in the event. However, some events may not contain all the required physics objects to reconstruct some features. In this case, the affected features in such events are padded with zero to ensure a consistent number of input features across the simulated datasets.

\begin{figure*}
    \centering
    \includegraphics[width=0.3\linewidth]{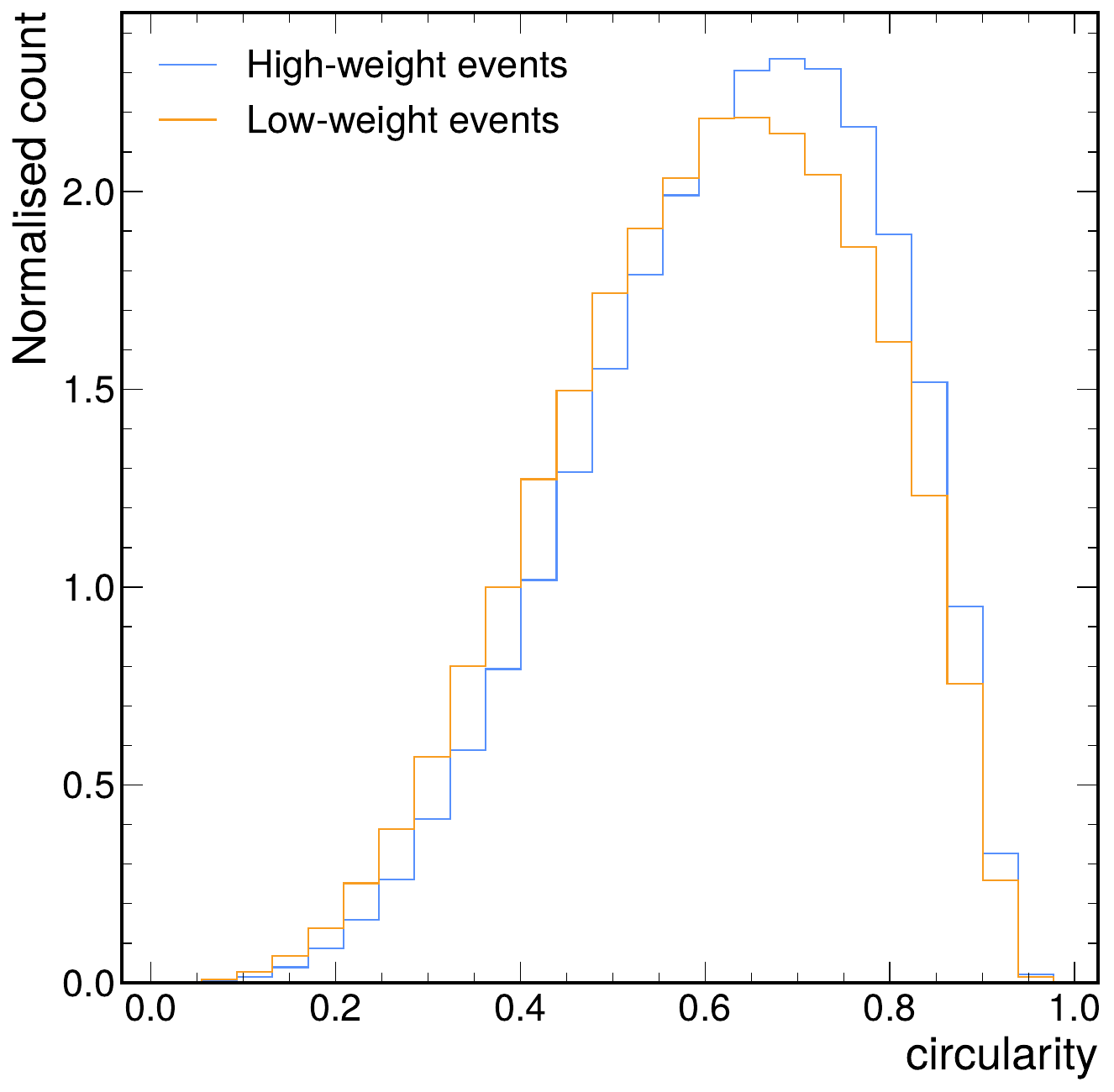}
    \includegraphics[width=0.3\linewidth]{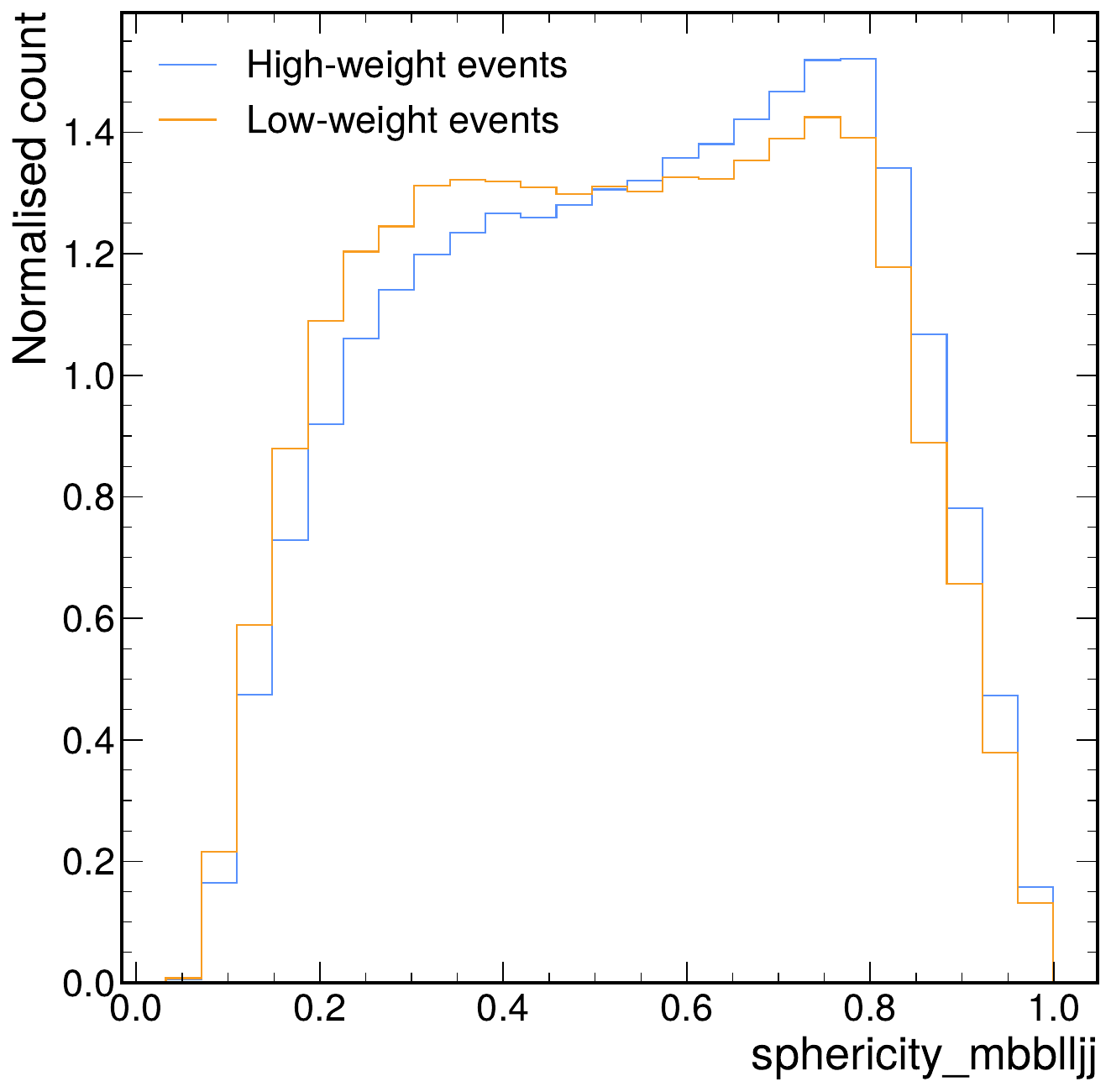}
    \includegraphics[width=0.3\linewidth]{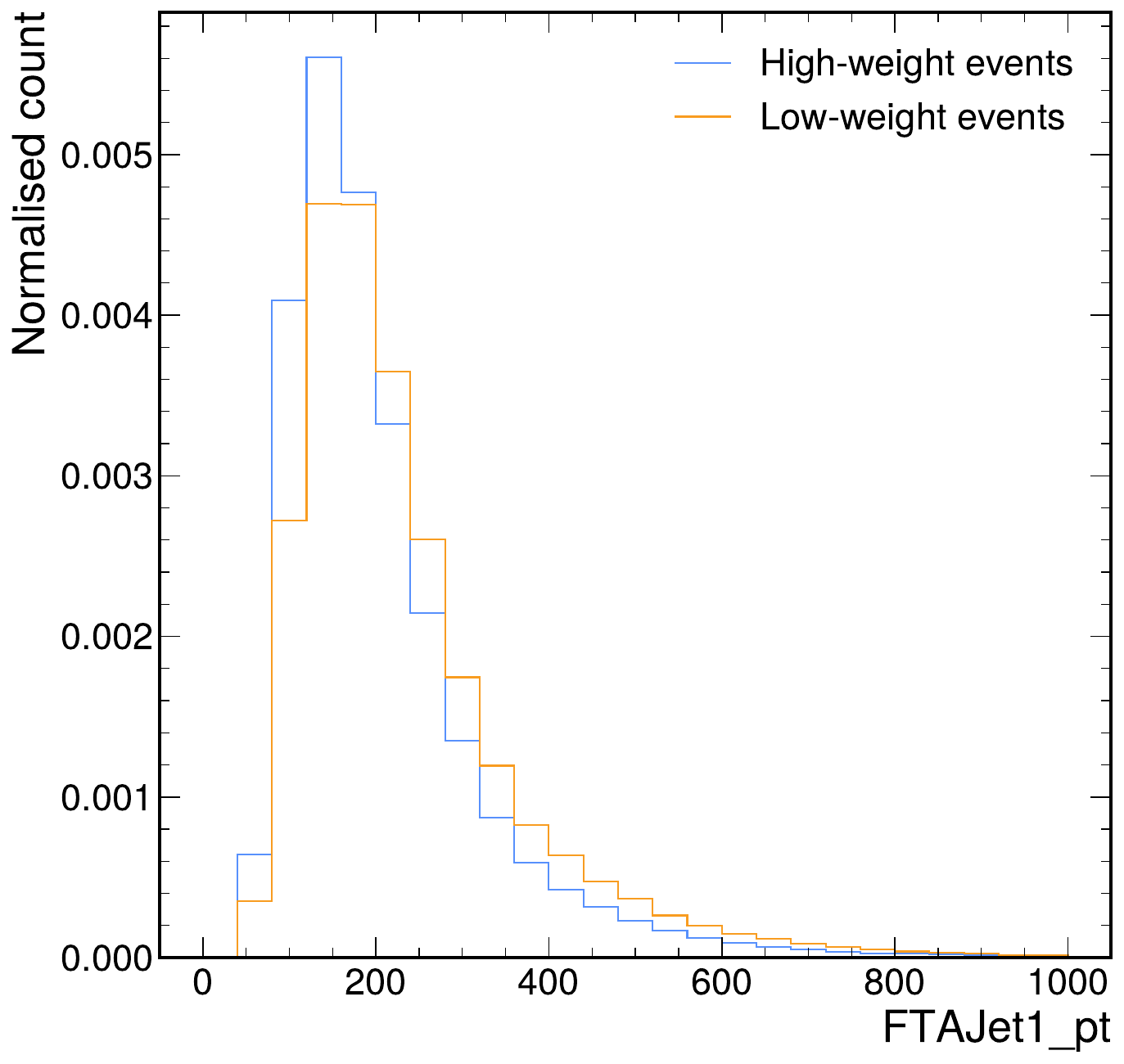}
    \caption{Example input features showing differences in the distribution between high-weight and low-weight events}
    \label{fig:input-features-example}
\end{figure*}

Initial observations towards the kinematic features and the event weights with respect to $y_t$ suggest that events with higher weight variance, where the event weight increases faster as $y_t$ increases, have certain kinematic feature distributions different from events with lower weight variance, where the event weight does not increase much as $y_t$ increases, compared to the former. Empirically, we can simply categorize the \fourtops events into these two groups using a criterion in the ratio between the event weight assigned at two different values of $y_t$, and use these two event categories to train our Machine Learning-based discriminators. For brevity, we will refer to events with a higher weight ratio as \emph{high-weight events}, while events with a lower weight ratio will be referred to as \emph{low-weight events}. Example distributions from three input features with this categorization rule are shown in Figure~\ref{fig:input-features-example}. 

\section{Learning strategy and model training} \label{sec:learning-strategy}
To exploit extra kinematic features from \fourtops production, we employ a two-stage learning strategy as follows:
\begin{itemize}
    %\item \textbf{NN\nobreakdash-c}, designed to classify events from \fourtops production from the two dominant backgrounds: \ttbar and $t\bar{t}H$ productions. This network is designed for maximum purity of events originated from \fourtops production.
    \item \textbf{Background rejection network} classifies events from \fourtops processes from the two dominant background processes: \ttbar and \ttH processes. This network is designed to enhance signal purity in the presence of both backgrounds, which is crucial given that the two dominant background processes have a much larger probability to occur inside the particle detectors.
    %\item \textbf{NN\nobreakdash-$y_t$}, designed to give different output distributions for \fourtops process with varying values of $y_t$.
    \item \textbf{Parameter inference network} learns parameter-sensitive features and returns one output representing the event's sensitivity to changes induced in the parameter of interest $y_t$. 
\end{itemize}

During parameter inference, an event will have outputs from both neural networks and will be used for the construction of template histograms, similar to the approach introduced in Ref.~\cite{TOP-21-001}.

While there are several approaches to parameter inference relying on unbinned likelihoods~\cite{Brehmer_2018}, where the probable parameter regions are inferred without using summary statistics in the form of histogram bins, we will utilize the binned approach, where histograms are constructed to represent summary statistics. The rationale for this approach is that it is simpler to implement and closely resembles the traditional parameter inference method, which requires histograms containing information on process yields as well. Hence, outputs from both networks will be used to categorize the events into histogram bins, which are optimized for both signal and background separation, and maximize yield differences with respect to the parameter of interest.

For brevity, the detailed architectures for both networks will be discussed in Appendix~\ref{app:nn-techincal-details}.

\subsection{Background rejection network} \label{sec:classification-network}
In the first part of the inference, the background rejection network is designed to distinguish the \fourtops events from the two dominant backgrounds. As both \ttbar and \ttH backgrounds have the largest contribution to the background, it is crucial for us to extract \fourtops events into a region that is least affected by the two dominant backgrounds. Based on this goal, the network will have three output nodes, each representing the $t\bar{t}t\bar{t}$, $t\bar{t}$, and \ttH processes.

%56 kinematic features are used as inputs to NN\nobreakdash-c, which are chosen based on the differences of the distributions of these features between $t\bar{t}t\bar{t}$, \ttbar, and \ttH processes. The features chosen for this network are calculated from final particles observable in particle detectors.

%The overall network structure contains an input layer accepting a set of 56 input features, followed by a batch normalisation layer and a number of feed-forward dense layer. Each feed-forward dense layer uses ReLU activation function, and is followed by a dropout layer with dropout probability of 0.1. The set of feed-forward dense layers is then followed by an output layer containing three nodes, representing $t\bar{t}t\bar{t}$, $t\bar{t}$, and \ttH processes. This output layer uses softmax activation function, which guarantees that the sum of the three output nodes will be 1.

\begin{figure}
    \centering
    \begin{tikzpicture}
        \filldraw[color=black, fill=red!75, thick] (0, 0) rectangle (0.5, 0.5);
        \filldraw[color=black, fill=red!75, thick] (0, 0.5) rectangle (0.5, 1.);
        \filldraw[color=black, fill=red!75, thick] (0, 1) rectangle (0.5, 1.5);
        \filldraw[color=black, fill=red!75, thick] (0, 1.5) rectangle (0.5, 2.);
        \filldraw[color=black, fill=red!75, thick] (0, 2) rectangle (0.5, 2.5);
        \filldraw[color=black, fill=red!75, thick] (0, 2.5) rectangle (0.5, 3.);
        \filldraw[color=black, fill=red!75, thick] (0, 3) rectangle (0.5, 3.5);
        \filldraw[color=black, fill=red!75, thick] (0, 3.5) rectangle (0.5, 4.);
        \filldraw[color=black, fill=red!75, thick] (0, 4) rectangle (0.5, 4.5);
        \filldraw[color=black, fill=red!75, thick] (0, 4.5) rectangle (0.5, 5.);
        \filldraw[color=black, fill=red!75, thick] (0.5, 0) rectangle (1., 0.5);
        \filldraw[color=black, fill=red!75, thick] (0.5, 0.5) rectangle (1., 1.);
        \filldraw[color=black, fill=red!75, thick] (0.5, 1) rectangle (1., 1.5);
        \filldraw[color=black, fill=red!75, thick] (0.5, 1.5) rectangle (1., 2.);
        \filldraw[color=black, fill=red!75, thick] (0.5, 2) rectangle (1., 2.5);
        \filldraw[color=black, fill=red!75, thick] (0.5, 2.5) rectangle (1., 3.);
        \filldraw[color=black, fill=red!75, thick] (0.5, 3) rectangle (1., 3.5);
        \filldraw[color=black, fill=red!75, thick] (0.5, 3.5) rectangle (1., 4.);
        \filldraw[color=black, fill=red!75, thick] (0.5, 4) rectangle (1., 4.5);
        \filldraw[color=black, fill=red!75, thick] (1, 0) rectangle (1.5, 0.5);
        \filldraw[color=black, fill=red!75, thick] (1, 0.5) rectangle (1.5, 1.);
        \filldraw[color=black, fill=red!75, thick] (1, 1) rectangle (1.5, 1.5);
        \filldraw[color=black, fill=red!75, thick] (1, 1.5) rectangle (1.5, 2.);
        \filldraw[color=black, fill=red!75, thick] (1, 2) rectangle (1.5, 2.5);
        \filldraw[color=black, fill=red!75, thick] (1, 2.5) rectangle (1.5, 3.);
        \filldraw[color=black, fill=red!75, thick] (1, 3) rectangle (1.5, 3.5);
        \filldraw[color=black, fill=red!75, thick] (1, 3.5) rectangle (1.5, 4.);
        \filldraw[color=black, fill=red!75, thick] (1.5, 0) rectangle (2., 0.5);
        \filldraw[color=black, fill=red!75, thick] (1.5, 0.5) rectangle (2., 1.);
        \filldraw[color=black, fill=red!75, thick] (1.5, 1) rectangle (2., 1.5);
        \filldraw[color=black, fill=red!75, thick] (1.5, 1.5) rectangle (2., 2.);
        \filldraw[color=black, fill=red!75, thick] (1.5, 2) rectangle (2., 2.5);
        \filldraw[color=black, fill=red!75, thick] (1.5, 2.5) rectangle (2., 3.);
        \filldraw[color=black, fill=red!75, thick] (1.5, 3) rectangle (2., 3.5);
        \filldraw[color=black, fill=red!75, thick] (2, 0) rectangle (2.5, 0.5);
        \filldraw[color=black, fill=red!75, thick] (2, 0.5) rectangle (2.5, 1.);
        \filldraw[color=black, fill=red!75, thick] (2, 1) rectangle (2.5, 1.5);
        \filldraw[color=black, fill=red!75, thick] (2, 1.5) rectangle (2.5, 2.);
        \filldraw[color=black, fill=red!75, thick] (2, 2) rectangle (2.5, 2.5);
        \filldraw[color=black, fill=red!75, thick] (2, 2.5) rectangle (2.5, 3.);
        \filldraw[color=black, fill=red!75, thick] (2.5, 0) rectangle (3., 0.5);
        \filldraw[color=black, fill=red!75, thick] (2.5, 0.5) rectangle (3., 1.);
        \filldraw[color=black, fill=red!75, thick] (2.5, 1) rectangle (3., 1.5);
        \filldraw[color=black, fill=red!75, thick] (2.5, 1.5) rectangle (3., 2.);
        \filldraw[color=black, fill=red!75, thick] (2.5, 2) rectangle (3., 2.5);
        \filldraw[color=black, fill=red!75, thick] (3, 0) rectangle (3.5, 0.5);
        \filldraw[color=black, fill=red!75, thick] (3, 0.5) rectangle (3.5, 1.);
        \filldraw[color=black, fill=red!75, thick] (3, 1) rectangle (3.5, 1.5);
        \filldraw[color=black, fill=red!75, thick] (3, 1.5) rectangle (3.5, 2.);
        \filldraw[color=black, fill=red!75, thick] (3.5, 0) rectangle (4., 0.5);
        \filldraw[color=black, fill=red!75, thick] (3.5, 0.5) rectangle (4., 1.);
        \filldraw[color=black, fill=red!75, thick] (3.5, 1) rectangle (4., 1.5);
        \filldraw[color=black, fill=red!75, thick] (4, 0) rectangle (4.5, 0.5);
        \filldraw[color=black, fill=red!75, thick] (4, 0.5) rectangle (4.5, 1.);
        \filldraw[color=black, fill=red!75, thick] (4.5, 0) rectangle (5., 0.5);

        \draw[very thick] (0, 0) node[anchor=north east] {0} -- (5, 0) node[anchor=north] {1.0} -> (5.5, 0) node[anchor=west]{\ttH};
        \draw[very thick] (0, 0) -- (0, 5) node[anchor=east] {1.0} -> (0, 5.5) node[anchor=south]{\fourtops};
        
    \end{tikzpicture}
    \caption{55 event categories separated by the output of the background rejection network. The $x$ and $y$-axes depict the output from \ttH and \fourtops nodes, respectively.}
    \label{fig:classification-network-category}
\end{figure}
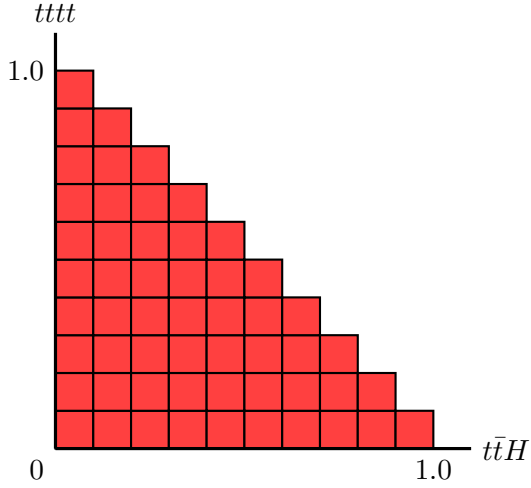

Hyperparameter tuning is used to determine the optimal number of layers and the number of neurons per layer. The goals of this tuning are to preserve the purity of \fourtops events as much as possible, while at the same time providing the continuous output distribution for each output node. The latter requirement will be used to ensure that at least one event must be present in each of the 55 event categories based on the \fourtops and \ttH output nodes, as shown in Figure~\ref{fig:classification-network-category}. Based on this categorization, events from $t\bar{t}t\bar{t}$, $t\bar{t}$, and \ttH processes should reside in one of the three corners of the distribution shown in the figure.

%Hyperparameter tuning is used to determine the optimal number of layers and the number of neurons per each layer. The goals of this tuning are to give the highest amount of \fourtops signal events versus both background events when the cut of 0.6 at the \fourtops output node is applied, while at the same time provide the continuous output distribution spanning in the range of $[0, 1]$.

%To ensure that the output of all nodes are continuous, events from both training and testing datasets are populated based on the output of \fourtops and \ttH nodes. Since the sum of all the outputs in each event must be 1 due to the softmax activation function in the output layer, all events are separated into 55 categories, depicted in Figure~\ref{fig:classification-network-category}. Each category spans the output range (from \fourtops and \ttH nodes) of 0.1. For instance, the $t\bar{t}t\bar{t}$-rich region, shown at the top left part of the figure, must have the \fourtops output between $[0.9, 1.0)$ and \ttH output of $[0.0, 0.1)$. During hyperparameter training, a network configuration will be vetoed if there is at least one out of 55 event categories where no events are populated. The same 55 event categories will also be used to construct the template histograms for $y_t$ inference later.

%Training datasets for this network contain 150\,000 events from each process to ensure equal representation during the training, and the remaining events for each process are put into testing datasets. The network is trained using categorical cross-entropy loss and Adam optimiser with the learning rate of 0.05. To prevent overtraining, early stopping criteria is applied where the training will stop if there are no improvements over the validation loss for five consecutive epochs, and the model with the best validation loss is saved during the training.

%The output distribution of the best network configuration, which contains three layers of 50, 100, and 100 neurons and dropout probability of 0.1, is shown in Figure~\ref{fig:output-dist-classification-model}. As seen from the figure, the output distributions span a complete output range between 0 and 1. This allows events from \fourtops signal and both backgrounds to be included in the maximum number of event categories. Figure~\ref{fig:output-dist-classification-model} also shows the distributions for each processes between training and testing datasets, showing no signs of overtraining. The model used in the later stages of this work is captured from the best training epoch during the training.

\begin{figure*}
    \centering
    \includegraphics[width=\linewidth]{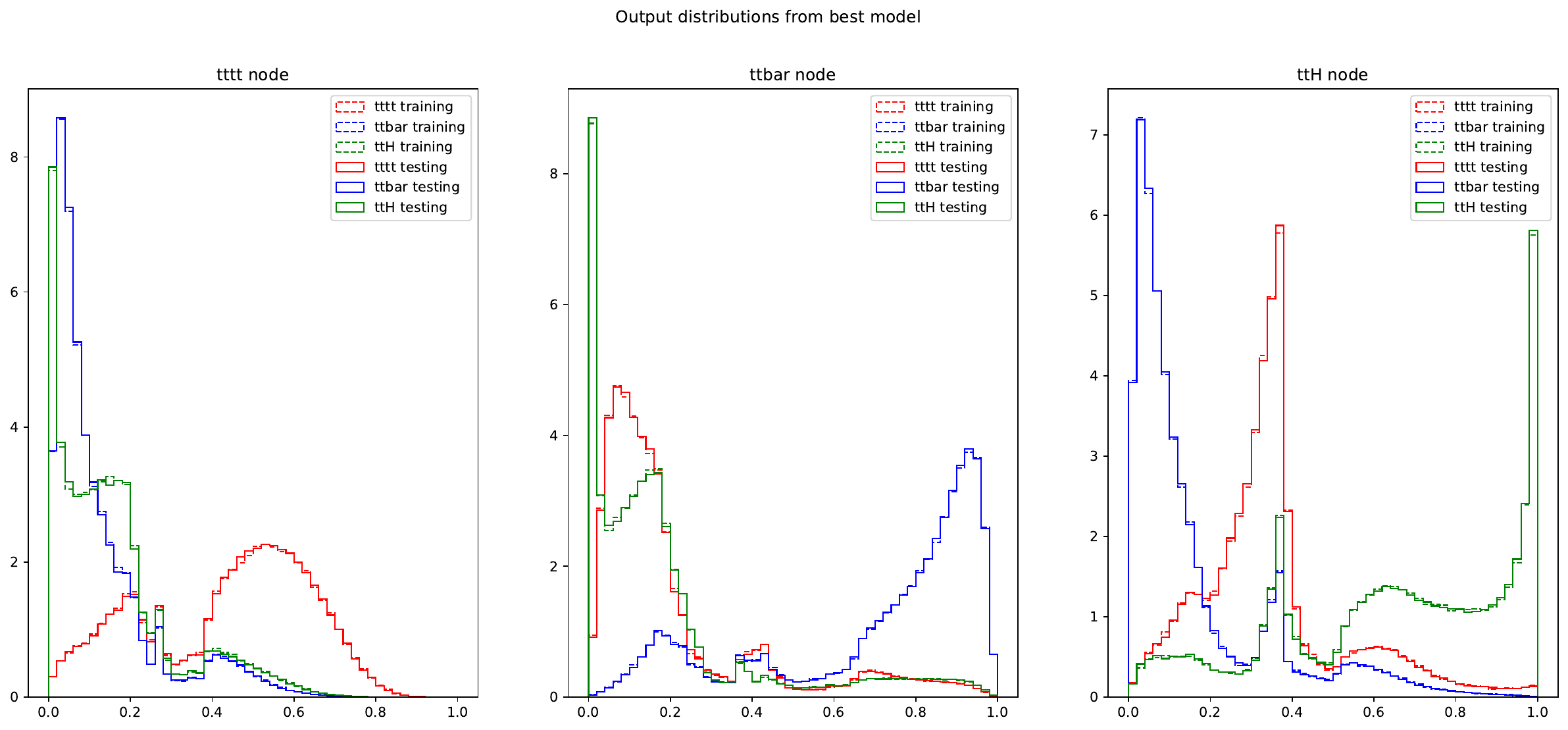}
    \caption{Output distribution of the best background rejection model according to the hyperparameter tuning, normalized. The distributions also do not show overtraining, as both distributions from the training (dashed lines) and testing (solid lines) datasets are comparable to each other.}
    \label{fig:output-dist-classification-model}
\end{figure*}

The output distribution of the best network configuration is shown in Figure~\ref{fig:output-dist-classification-model}. As seen from the figure, the output distributions span a complete output range between 0 and 1. This allows events from \fourtops signal and both backgrounds to be included in the maximum number of event categories. Figure~\ref{fig:output-dist-classification-model} also shows the distributions for each process between training and testing datasets, showing no signs of overtraining. The model used in the later stages of this work is captured from the best training epoch during the training.

Technical details for the training of this network are available in Appendix~\ref{subapp:bg-rejection-network-details}.

\subsection{Parameter inference network} \label{subsection:yt-network}
In the second part of inference, the parameter inference network is designed to give different output distributions for the \fourtops process as the $y_t$ parameter value changes. To achieve this, the network must discriminate input events between low-weight and high-weight categories. With this design, as $y_t$ changes, the network output distribution will be skewed towards the regions with more high-weight events. 

The rationale behind this design is that, as \fourtops events are sensitive to the $y_t$ parameter, their event weights should reflect this. However, as observed in Section~\ref{subsec:feature-representation}, not all \fourtops events have a high variance in event weights at different values of sampled $y_t$. If we can discriminate \fourtops events into two groups, low-weight events (with low weight variance) and high-weight events (with high weight variance), we can have a distribution shape based on the output of this discriminator that \emph{skews} per the change in $y_t$, therefore being sensitive to the parameter of interest.

Hyperparameter tuning is also applied to this network, albeit with a different objective. Since the goal of this network is to discriminate between both low-weight and high-weight events and obtain the distribution that is sensitive to the $y_t$ parameter, the procedure is aimed at determining the highest minimum change ratio compared between two points of $y_t$ values. The best network determined by this tuning approach should provide significant differences between $y_t$ values for event yields regardless of the histogram bins.

%The training dataset for this network contains 250\,000 \fourtops events from each of the two event categories assigned by the weight ratio between two points of $y_t$ values, while the remaining events are gathered as the testing dataset. This network is trained using binary cross-entropy loss and Adam optimiser with the learning rate of 0.05. The same early stopping criteria is also applied where the training will stop after three epochs of no improvement over the validation loss.

%The optimal network from hyperparameter tuning contains four layers of 200, 200, 50, and 50 neurons, and the dropout probability of 0.2. Figure~\ref{fig:output-dist-train-test} illustrates the training output distribution from the optimal $y_t$ network, as determined by the hyperparameter tuning requirement. The figure also shows no signs of overtraining as the output distributions between the training and testing datasets are similar.

\begin{figure}
    \centering
    \includegraphics[width=0.5\linewidth]{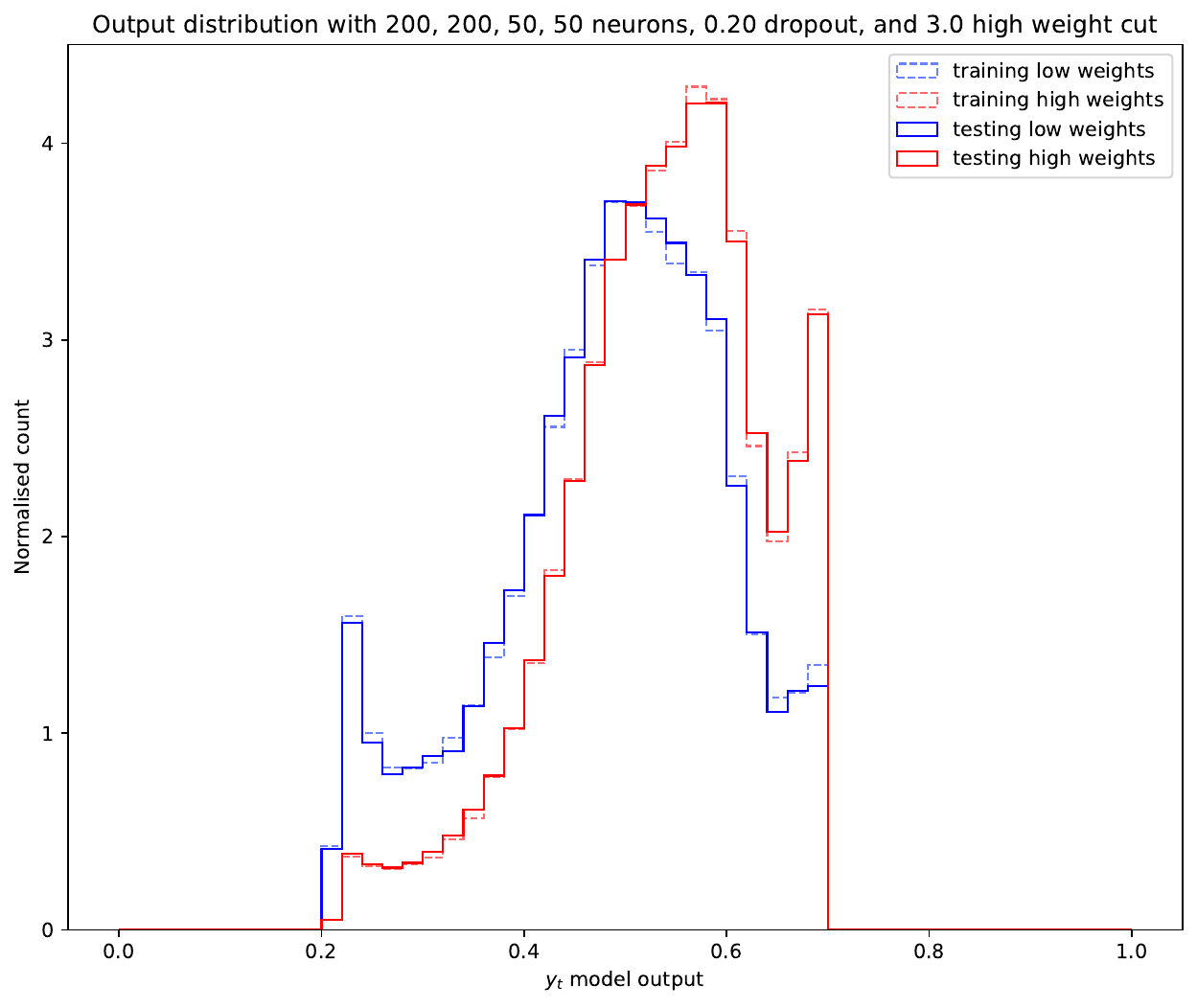}
    \caption{Output distribution of the best parameter inference network determined by hyperparameter tuning, showing the differences between high-weight events versus low-weight events. The distributions also do not show overtraining, as both distributions from the training (dashed lines) and testing (solid lines) datasets are comparable to each other.}
    \label{fig:output-dist-train-test}
\end{figure}

Figure~\ref{fig:output-dist-train-test} illustrates the training output distribution from the optimal $y_t$ network, as determined by the hyperparameter tuning requirement. No significant discrepancies between training and testing output distributions are observed.
%shows no signs of overtraining as the output distributions between the training and testing datasets are similar.

\begin{figure*}
    \centering
    \begin{subfigure}[b]{0.75\linewidth}
         \centering
        \includegraphics[width=\linewidth]{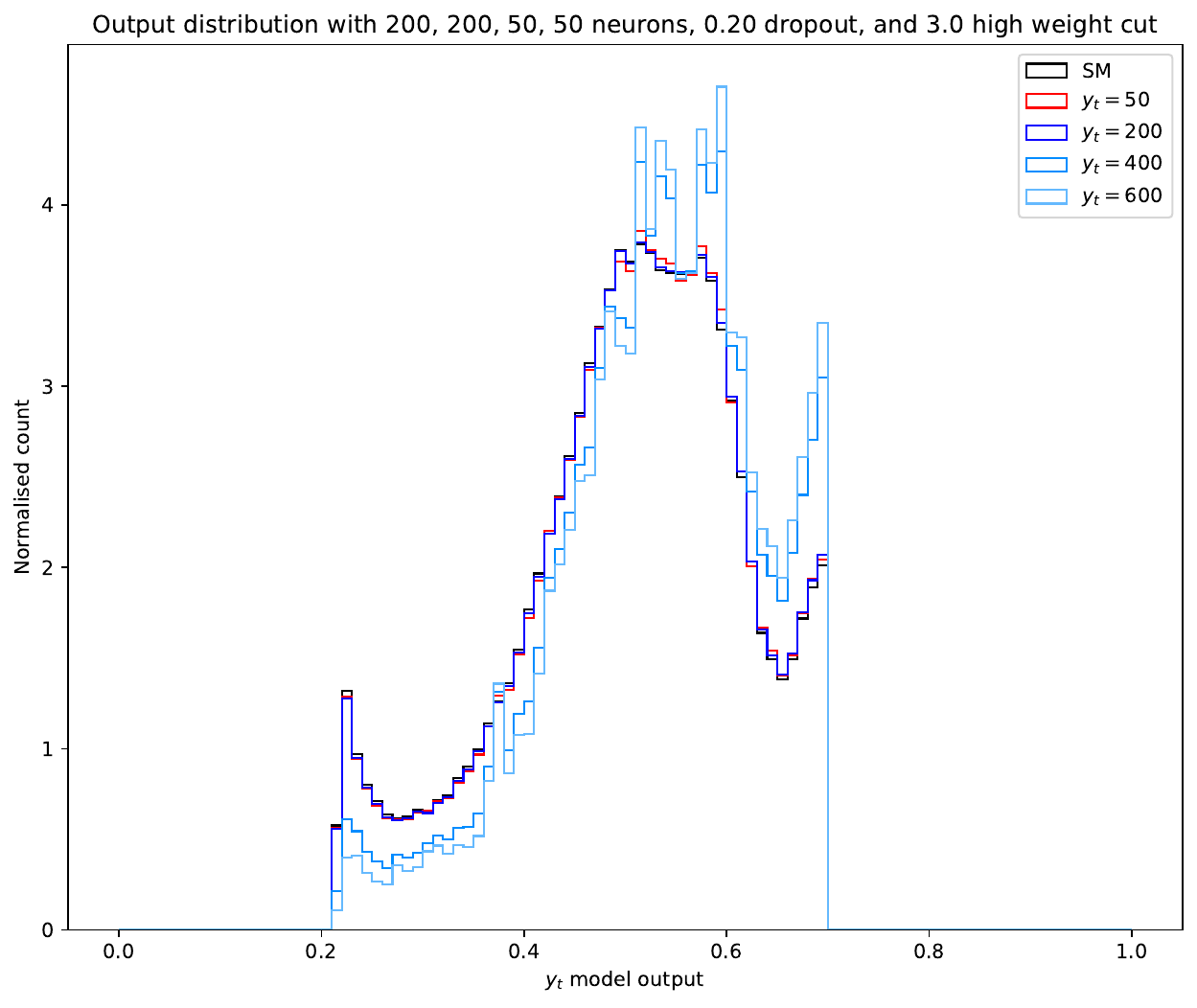}
         \caption{with $y_t$ weights applied}
         \label{fig:output-distribution-yt}
    \end{subfigure}
    \centering
    \begin{subfigure}[b]{0.45\linewidth}
         \centering
        \includegraphics[height=5cm]{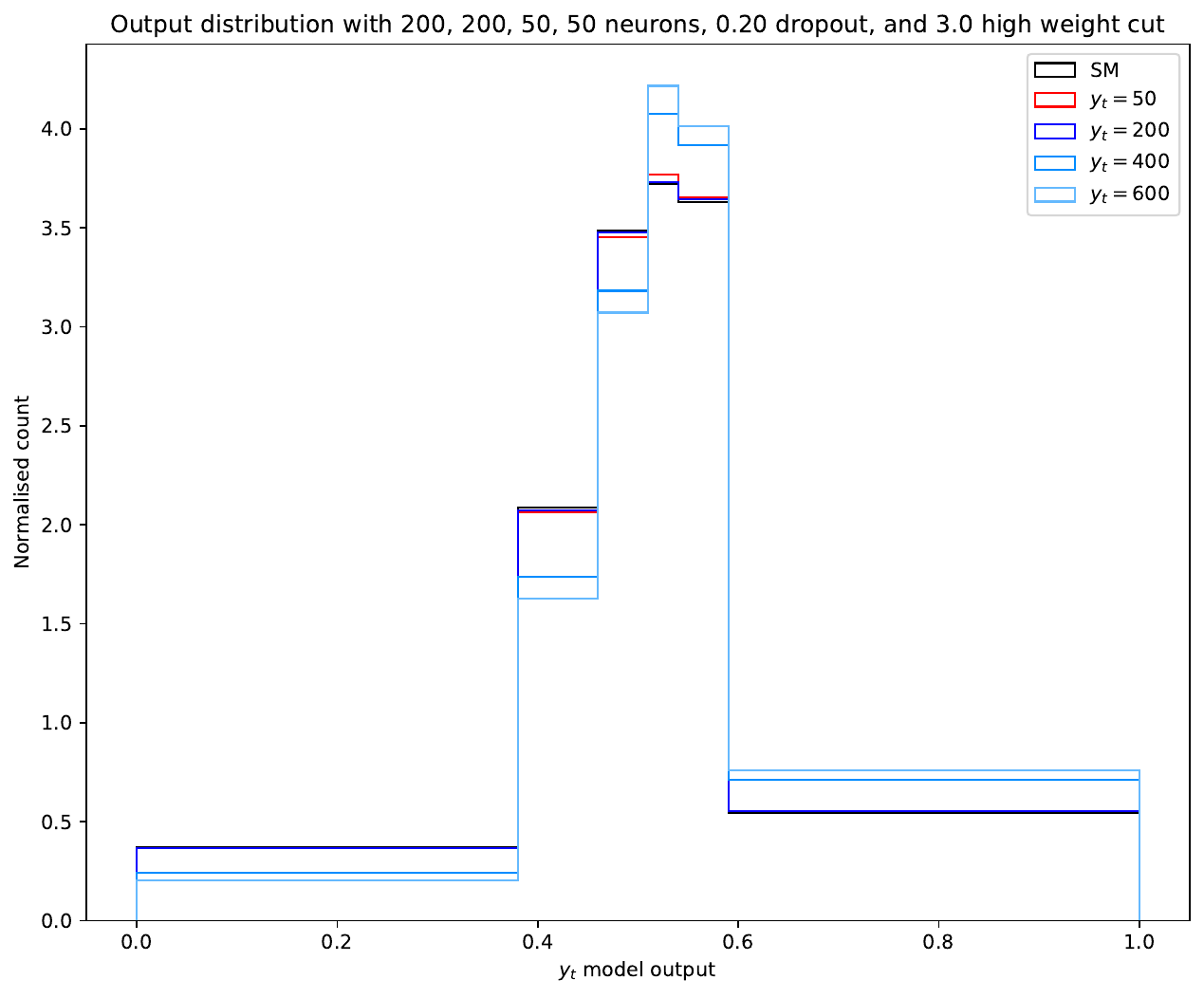}
         \caption{with optimal binning}
         \label{fig:output-distribution-yt-optimal-cuts}
    \end{subfigure}
    \centering
    \begin{subfigure}[b]{0.45\linewidth}
        \centering
        \includegraphics[height=5cm]{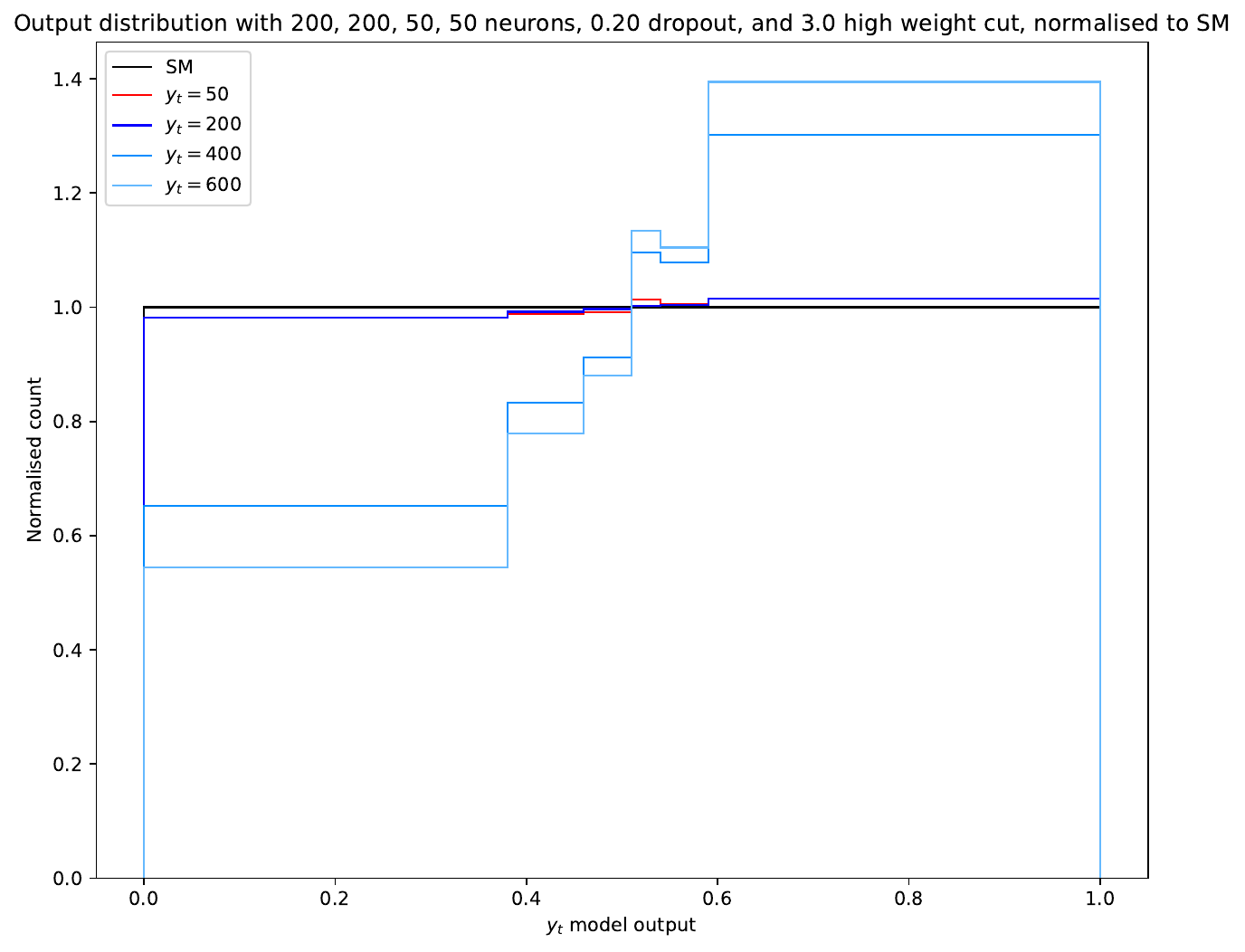}
        \caption{with optimal binning, normalized around SM yields}
        \label{fig:output-distribution-yt-optimal-cuts-normalised}
    \end{subfigure}
    \caption{Optimal binning from the best parameter inference network per hyperparameter tuning. Figure~\ref{fig:output-distribution-yt} shows the output distribution from \fourtops events with different event weights per $y_t$ applied, where each histogram is normalized to its yield in order to highlight the shape difference induced by the change in $y_t$. The same distribution is then put into an optimal binning of six bins, resulting in Figure~\ref{fig:output-distribution-yt-optimal-cuts} To determine the best configuration for this network, the histogram in Figure~\ref{fig:output-distribution-yt-optimal-cuts} is normalized, where the yields at the default value of $y_t$ parameter as dictated by SM are set to 1, resulting in the histogram in Figure~\ref{fig:output-distribution-yt-optimal-cuts-normalised}.}
\end{figure*}

As the network is supposed to give higher output values to high-weight events, the overall distribution \emph{shape} will skew towards the region with higher output value as $y_t$ increases. This effect is illustrated in Figure~\ref{fig:output-distribution-yt}. The final binning, as determined during the training, is shown in Figure~\ref{fig:output-distribution-yt-optimal-cuts}, and the ratio between the event yield at different $y_t$ values versus the default value as dictated by SM is shown in Figure~\ref{fig:output-distribution-yt-optimal-cuts-normalised}. The final binning shown in Figure~\ref{fig:output-distribution-yt-optimal-cuts} will be used along with the best version of the network with this configuration.

Technical details for the training of this network are available in Appendix~\ref{subapp:param-inference-network-details}.

\section{Parameter inference methodology} \label{sec:methodology}
The learning strategy explained in Section~\ref{sec:learning-strategy} results in two networks providing low-dimensional representations that must be used both for background rejection and $y_t$ parameter inference. The representation for the latter goal is distilled by the parameter inference network as one-dimensional and should provide a distribution that varies smoothly with the parameter $y_t$. In this section, we will construct the well-defined summary statistics based on these representations and use them in a likelihood-based parameter inference in two ways: direct inference from the constructed statistics and indirect inference from a surrogate quantity.

\subsection{Template histogram construction} \label{subsec:template-histogram-construction}
Based on the output of the background rejection network, events from $t\bar{t}t\bar{t}$, $t\bar{t}$, and \ttH are categorized into the same 55 event categories as described in Section~\ref{sec:classification-network} and depicted in Figure~\ref{fig:classification-network-category}. The background rejection output should guarantee that all 55 event categories have a nonzero number of events in each of them.

In each event category, the events are then further separated by the output of the parameter inference network into summary statistics in the form of a histogram with six bins. The binning for each histogram is already determined in Section~\ref{subsection:yt-network}. However, if the binning results in one or more empty histogram bins, those empty histogram bins are then combined.

%The final result of this categorisation is 55 template histograms, each containing at least one histogram bin, and at most six histogram bins. They are then scaled by their cross section values per each process, both signal and backgrounds. They are scaled such that the number of events reflect the actual amount of events for any given amount of integrated luminosity. Furthermore, this work will mainly focus on the three integrated luminosity points, which are 49.81 fb$^{-1}$ (2017 amount of data from CMS), 138 fb$^{-1}$ (2016--2018 amount of data from CMS), and 3000 fb$^{-1}$ (expected amount of data attainable from High-Luminosity LHC upgrade, HL-LHC).

The final result of this categorization is 55 histograms, each containing at least one histogram bin and at most six histogram bins. They are then scaled to reflect the total yields for each process at certain amounts of data obtainable from the particle detector. Specifically, we will focus on three levels of detector data amounts:
\begin{itemize}
    \item Amount of data gathered from CMS detector in 2017
    \item Amount of data gathered from CMS detector in 2016--2018 (Full Run 2)
    \item Expected amount of data attainable from High-Luminosity LHC
    %\item Expected amount of data attainable from Large Hadron Collider (LHC) with High-Luminosity LHC upgrade (HL-LHC)
\end{itemize}

\begin{figure*}
    \centering
    \includegraphics[width=\linewidth]{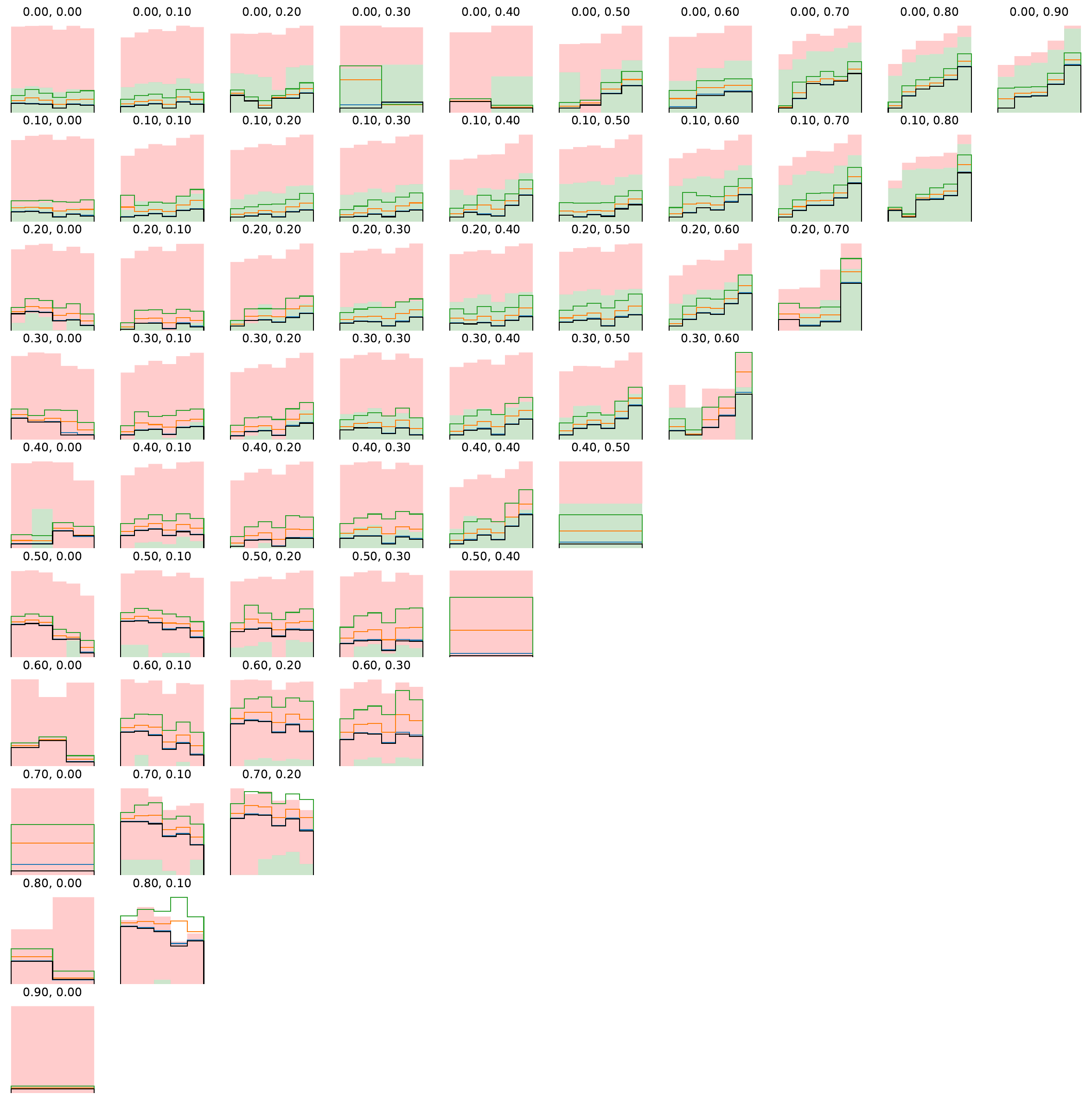}
    \caption{Template histograms used for $y_t$ inference, separated by the categories per classification network output. The label at each histogram shows the starting value of \fourtops and \ttH output node, respectively. Each category covers the range of \fourtops and \ttH output node values of 0.1 each. Red solid histograms represent $t\bar{t}$, green solid histograms represent $t\bar{t}H$, and the black line represents \fourtops distributions at the nominal $y_t$ parameter value assigned in SM. Blue, orange, and green lines represent \fourtops distributions with non-nominal values of $y_t$. $y$-axes in all histograms are in log scale.}
    \label{fig:fitting-mountainrange-meta}
\end{figure*}

\begin{figure*}
    \centering
    \includegraphics[width=\linewidth]{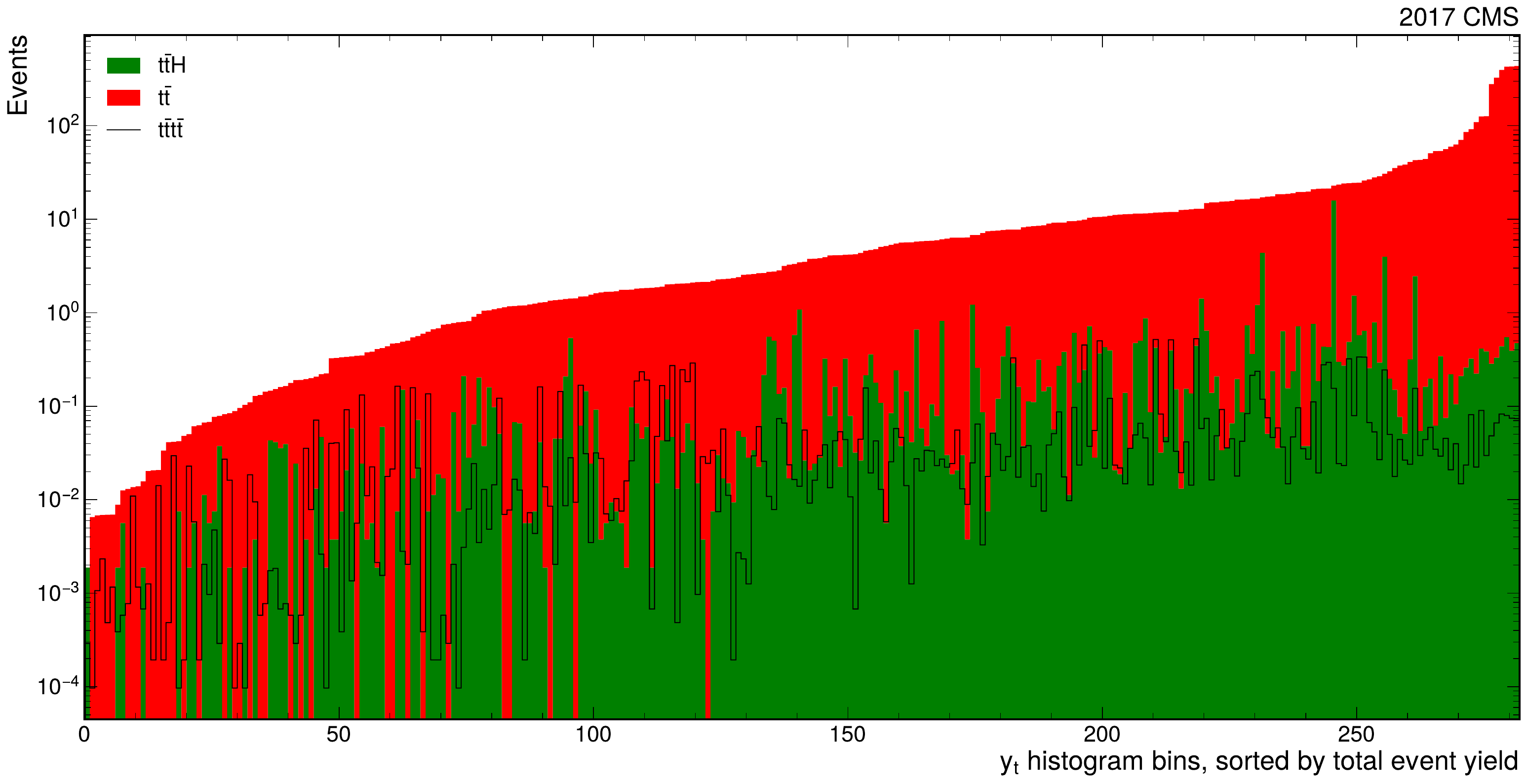}
    \caption{Event yields from histogram bins shown in Figure~\ref{fig:fitting-mountainrange-meta}, sorted by total background event yield and normalized to the 2017 CMS data amount.}
    \label{fig:fitting-mountainrange-meta-yield}
\end{figure*}

The final template histograms are shown in Figures~\ref{fig:fitting-mountainrange-meta} and~\ref{fig:fitting-mountainrange-meta-yield}. As shown in Figure~\ref{fig:fitting-mountainrange-meta}, the template histograms contain the $t\bar{t}H$-rich region where most of the events have high \ttH node output. There is also the $t\bar{t}t\bar{t}$-rich region where almost no \ttH events are present, at the expense of low statistics, causing the histograms to contain a very small number of bins. To compare the event yields between different histogram bins represented in Figure~\ref{fig:fitting-mountainrange-meta}, they are rearranged in Figure~\ref{fig:fitting-mountainrange-meta-yield} by the total background event yield. Event yields at 2017 CMS data amount from the template histograms are shown in Figure~\ref{fig:fitting-mountainrange-meta-yield}. However, rearranging histogram bins in this way does not affect the parameter inference in any way.

\subsection{Direct parameter inference} \label{subsec:direct-yt-inference}
The template histograms constructed in Section~\ref{subsec:template-histogram-construction} can then be directly used for parameter inference. Given that we have already obtained event weights sampled from certain values of $y_t$ picked from a continuous range, we can parametrize the yields of each process, $t\bar{t}t\bar{t}$, $t\bar{t}$, and $t\bar{t}H$, such that they can be written as a function of the parameter. This allows us to estimate the yields for all processes freely within the continuous range of the parameter and allows us to infer the parameter value without being constrained to discrete values from parameter values chosen for simulation sampling.

For simplicity, we can parametrize the yields of each process in each histogram bin by the ratio of top Yukawa coupling to the nominal value by SM $Y_t = |y_t / y_t^\textrm{SM}|$ as follows:
\begin{enumerate}
    %\item $t\bar{t}t\bar{t}$: Since \fourtops process has the tree-level diagram containing two top-Higgs vertices, the event yield for this process per each histogram bin can be fitted to a 4th-order polynomial with respect to $Y_t$. In other words,
    \item $t\bar{t}t\bar{t}$: Yields for this process per each histogram bin can be fitted to a 4th-order polynomial with respect to $Y_t$. In other words,
    \begin{eqnarray}
        s_{t\bar{t}t\bar{t},\,i} (Y_t) &=& c_{0, i} + c_{1, i}Y_t + c_{2, i}Y_t^2 + c_{3, i}Y_t^3 + c_{4, i}Y_t^4
    \end{eqnarray}
    where $s_{t\bar{t}t\bar{t},\,i}$ is the \fourtops event yield in $i$-th histogram bin in the fitting distribution, and $c_{j, i}$ is a $j$-th order coefficient for the fitting polynomial. This means that each histogram bin $i$ will have a unique fourth-order polynomial describing the yield of \fourtops with respect to $Y_t$.
    The polynomial is fitted using the event yields calculated from event weights provided by the simulator~\cite{MadGraph5}. 
    
    This method of parametrizing yields for this process follows the motivation presented by Ref.~\cite{Cao_2019}, which shows that the cross section of this process can be scaled in the same fashion.
    %\item $t\bar{t}$: The yields for this process, dependent on $Y_t$, are varied by the electroweak corrections. It has been shown in Ref.~\cite{TOP-19-008} that the yield of this process can be fitted into a 2nd-order polynomial with respect to $Y_t$. Therefore, the yields of this process per each histogram bin are then fitted in the same way, using the event yields calculated from the electroweak corrections generated by Hathor software package~\cite{HATHOR}.
    \item $t\bar{t}$: Yields for this process can be fitted to a 2nd-order polynomial with respect to $Y_t$, in the same way as \fourtops process. The polynomial is fitted using a separate simulator calculating event weights as a correction for this process~\cite{HATHOR}. 
    
    The parameterization for this process follows the motivation presented by Ref.~\cite{TOP-19-008}, which shows that the yields for this process can be fitted to a second-order polynomial with respect to $Y_t$.
    \item $t\bar{t}H$: Yields for this process are always scaled up with $Y_t^2$ in all histogram bins.
\end{enumerate}

Based on this parametrization, we can construct a likelihood function 
\begin{equation}
    L(\textrm{data} | Y_t, \theta) = \prod_{i}\frac{\left(s_i(Y_t) + b_i(Y_t)\right)^{n_i}}{n_i!} e^{-(s_i(Y_t) + b_i(Y_t))} \label{eqn:likelihood-func}
\end{equation}
where $i$ represents a histogram bin in the fitting distribution, $s_i$, $b_i$, and $n_i$ represent the event yields for \fourtops signal events, \ttbar + \ttH background events, and actual data event yields from particle detectors, and $\theta$ represents uncertainties during the inference. The addition of uncertainties in this function allows us to take into account possible uncertainties that may arise during the parameter inference over actual data from particle detectors.

The likelihood function above is modified from the original function presented in Ref.~\cite{Cowan_2011}, where all the expected event yields $s_i$ and $b_i$ are parametrized as a function of $Y_t$. Since the yields for $t\bar{t}t\bar{t}$, $t\bar{t}$, and \ttH in each bin of each histogram have been parametrized as polynomials of $Y_t$, it is possible to evaluate the negative log-likelihood (NLL) $-2\ln L$ with respect to the parameter, in the same fashion as in Ref.~\cite{TOP-21-001}. %The evaluated NLL can be used to constrain the $Y_t$ value to the level of 68\% or even 95\% confidence level. The measurement is performed using Higgs Combine program from CMS Collaboration~\cite{higgs-combine}.

%As this is a proof-of-concept work, no real data from particle detectors is used. Instead, this work will present the expected measurement of the coupling based on Asimov dataset~\cite{Cowan_2011}. The Asimov dataset used for the measurement is constructed with the expected yields from the three processes, based on the MC samples, where $Y_t$ is set at the nominal SM value of 1. This means, for Equation~\ref{eqn:likelihood-func},
As this is a proof-of-concept work, no real data from particle detectors is used. Instead, this work will perform inference by assuming
\begin{equation}
    n_i = s_i(Y_t=1) + b_i(Y_t=1)
\end{equation}
As such, the inferred value of $Y_t$ will always be at $Y_t = 1$. However, the inference method still allows us to estimate the probable range of the parameter within certain levels of confidence. The parameter range inferred by the likelihood function can be affected by systematic uncertainties and statistical uncertainties (induced by low event yields).
%As such, the measured value obtained by this method will always have the nominal value of $Y_t = |y_t / y_t^\textrm{SM}| = 1$. This method still allows us to calculate measurement uncertainties, both systematic and statistical, which allows us to benchmark the measurement methods between different analyses. Lower measurement uncertainties indicate that the measurement method provides better accuracy.

Furthermore, the inference can also be done where the event yields for \ttH and \ttbar are not parametrized by $Y_t$. In this case, the \ttbar event yields are fixed to the nominal expected yield at $Y_t=1$ regardless of $Y_t$, and the \ttH yields are adjusted by a separate free parameter in the inference. As \ttH normalization becomes a free parameter, the entire yields of \ttH events may scale up or down without any constraints tied to the parameter of interest $Y_t$. In other words,
\begin{equation}
    b_i = b_{t\bar{t}, i}(Y_t = 1) + \mu_{\textrm{free}} \;b_{t\bar{t}H, i}
\end{equation}
where $\mu_\textrm{free}$ is the free parameter which is identical to all histogram bins.
This setting is similar to the inference additionally performed in Ref.~\cite{ATLAS-fourtops-observation}, where the yields for \ttH background are allowed to be scaled freely. The parameter inference with this setting can also be interpreted as a benchmark for our approach exploiting extra kinematic information from the \fourtops process.

%A minimal set of systematic uncertainties are included into $Y_t$ measurement in the form of nuisance parameters~\cite{Cowan_2011} as follows:
%\begin{enumerate}
%    \item \textbf{Luminosity} at 2\%~\cite{TOP-21-005}
%    \item \textbf{Jet energy scale} and \textbf{Jet energy resolution}, treated additionally with LOWESS smoothing algorithm~\cite{lowess-1,lowess-2}%, applied and correlated on all processes
%    \item \textbf{Renormalisation and factorisation}% ($\mu_R$ and $\mu_F$)%, applied but not correlated on all processes
%    \item \textbf{Initial and final state radiation}%, applied but not correlated on all processes
%    \item $\mathbf{t\bar{t} + b\bar{b}}$ \textbf{normalisation} at 8\%~\cite{TOP-21-005}
%    \item $\mathbf{t\bar{t}}$ \textbf{cross section} at 5.5\%~\cite{TOP-21-005}
%    \item $\mathbf{t\bar{t}H}$ \textbf{cross section} at 20\%~\cite{TOP-21-005}.
%    \item $\mathbf{t\bar{t}t\bar{t}}$ \textbf{cross section} at 18.7\%, calculated from the lower uncertainty of the \fourtops cross section at NLO (QCD+EW) + NLL' order of $14.65^{+1.57}_{-2.75}$~fb~\cite{vanbeekveld2025thresholdresummationproductionquarks}. %This uncertainty is not included in the \fourtops cross section measurement.
%\end{enumerate}

A minimal set of systematic uncertainties is assigned to the parameter inference, modeled after a physics analysis aimed at the \fourtops production search in the same final state in Ref.~\cite{TOP-21-005}.

%Since this work focuses on the \fourtops production with the final state comprising two opposite-sign leptons, the list of uncertainties above (except for \fourtops cross section uncertainty) is minimally modelled after a physics analysis aimed at the \fourtops production search in the same final state in Ref.~\cite{TOP-21-005}.

\subsection{Traditional inference via surrogate quantity} \label{subsec:template-histogram-xsect-measurement}
To benchmark the effectiveness of this weight-based inference method, we also performed the parameter inference in a traditional method, where a surrogate quantity, most commonly the cross section of a process, is inferred and then compared to the theoretical predictions of the same quantity with respect to the parameter.

In this work, the cross-section of \fourtops production is inferred using the same template histograms. Instead of adjusting yields with the $Y_t$ parameter, the yields of \fourtops production in all histogram bins are adjusted directly by the same factor called signal strength $\mu$, which allows us to derive the cross section based on the likelihood function~\cite{Cowan_2011}
\begin{equation}
    L(\textrm{data} | Y_t, \theta) = \prod_{i}\frac{\left(\mu s_i + b_i\right)^{n_i}}{n_i!} e^{-(\mu s_i + b_i)} \label{eqn:likelihood-func-xsect}
\end{equation}
The signal strength $\mu$ inferred from the template histograms can be simply translated to the inferred cross section by multiplying it by the \fourtops cross section of 14.65~fb~\cite{vanbeekveld2025thresholdresummationproductionquarks}, as the \fourtops yields in the template histograms are already adjusted with this amount of cross section. For indirect inference, we will assume that $\mu = 1$, leading to
\begin{equation}
    n_i = s_i + b_i
\end{equation}

Based on the \fourtops cross section prediction with respect to $Y_t$, adapted from Ref.~\cite{Cao_2019} with the SM cross section prediction from Ref.~\cite{vanbeekveld2025thresholdresummationproductionquarks},
\begin{eqnarray}
    \sigma(Y_t)  = \frac{7.724 - 1.164 Y_t^2 + 0.910 Y_t^4}{7.724 - 1.164 + 0.910} (14.65\textrm{ fb}) \label{eqn:fourtops-yield-yt}
\end{eqnarray}
we can then infer the probable range of $Y_t$ within a positive region by comparing the inferred cross section against the prediction in Equation~\ref{eqn:fourtops-yield-yt}. The inferred range of $Y_t$ can then be compared directly with the inferred range of $Y_t$ from the direct inference detailed in Section~\ref{subsec:direct-yt-inference}. This indirect inference method has been performed by both ATLAS and CMS experiments as an auxiliary interpretation in addition to the discovery of the \fourtops process~\cite{TOP-21-005,TOP-18-003,ATLAS-fourtops-observation}.

The same set of systematic uncertainties used in the direct inference is also included in this inference method, with slight adjustments to better suit the indirect inference via cross-section quantity.

\section{Results from parameter inference methods}
In this section, we will discuss the evaluation of the proposed method based on the inferred values of $Y_t$, compared to a traditional inference method via surrogate quantity.

\begin{figure}
    \centering
    \includegraphics[width=0.5\linewidth]{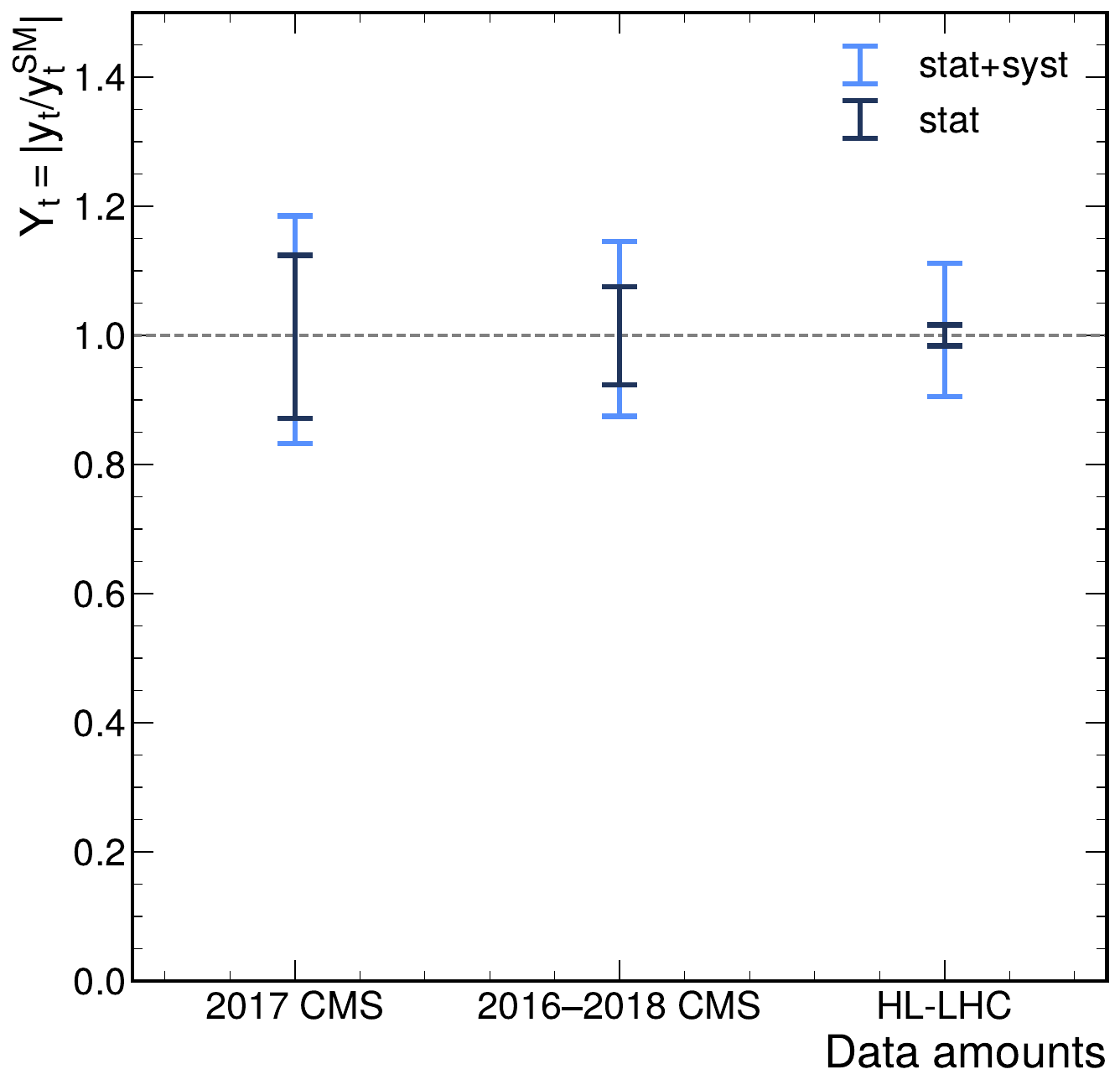}
    \caption{Expected inferred range of $Y_t = |y_t/y_t^\textrm{SM}|$ at different amounts of data.}
    \label{fig:yt-results}
\end{figure}

The inferred ranges of $Y_t = |y_t/y_t^\textrm{SM}|$, based on $Y_t$ parametrization over the template histograms at three different levels of data amounts, are shown in Table~\ref{tab:yt-results} and Figure~\ref{fig:yt-results}. As expected, the statistical uncertainties from the direct inference method decrease as the amount of data increases. However, systematic uncertainties do not decrease. This suggests that systematic uncertainties must be improved to increase the sensitivity of the coupling measurement.

\begin{table}
    \centering
    \caption{Expected inferred ranges of $Y_t = |y_t/y_t^\textrm{SM}|$ at different data amounts. The uncertainties listed are systematic and statistical uncertainties, respectively. The measurement is done at 68\% CL.}
    \label{tab:yt-results}
    \begin{tabular}{lcc}
        \hline
        & \multicolumn{2}{c}{$Y_t = |y_t/y_t^\textrm{SM}|$} \\
         Data amount    & syst. + stat. & combined \\ \hline
         2017 CMS       & $1^{+0.137}_{-0.108}\,^{+0.124}_{-0.129}$ & $1^{+0.185}_{-0.168}$ \\
         2016--2018 CMS & $1^{+0.124}_{-0.099}\,^{+0.075}_{-0.077}$ & $1^{+0.145}_{-0.125}$ \\
         HL-LHC         & $1^{+0.110}_{-0.093}\,^{+0.016}_{-0.016}$ & $1^{+0.112}_{-0.095}$\\ \hline
    \end{tabular}
\end{table}

\begin{figure*}
    \centering
    \begin{subfigure}[b]{0.45\linewidth}
        \centering
        \includegraphics[width=\textwidth]{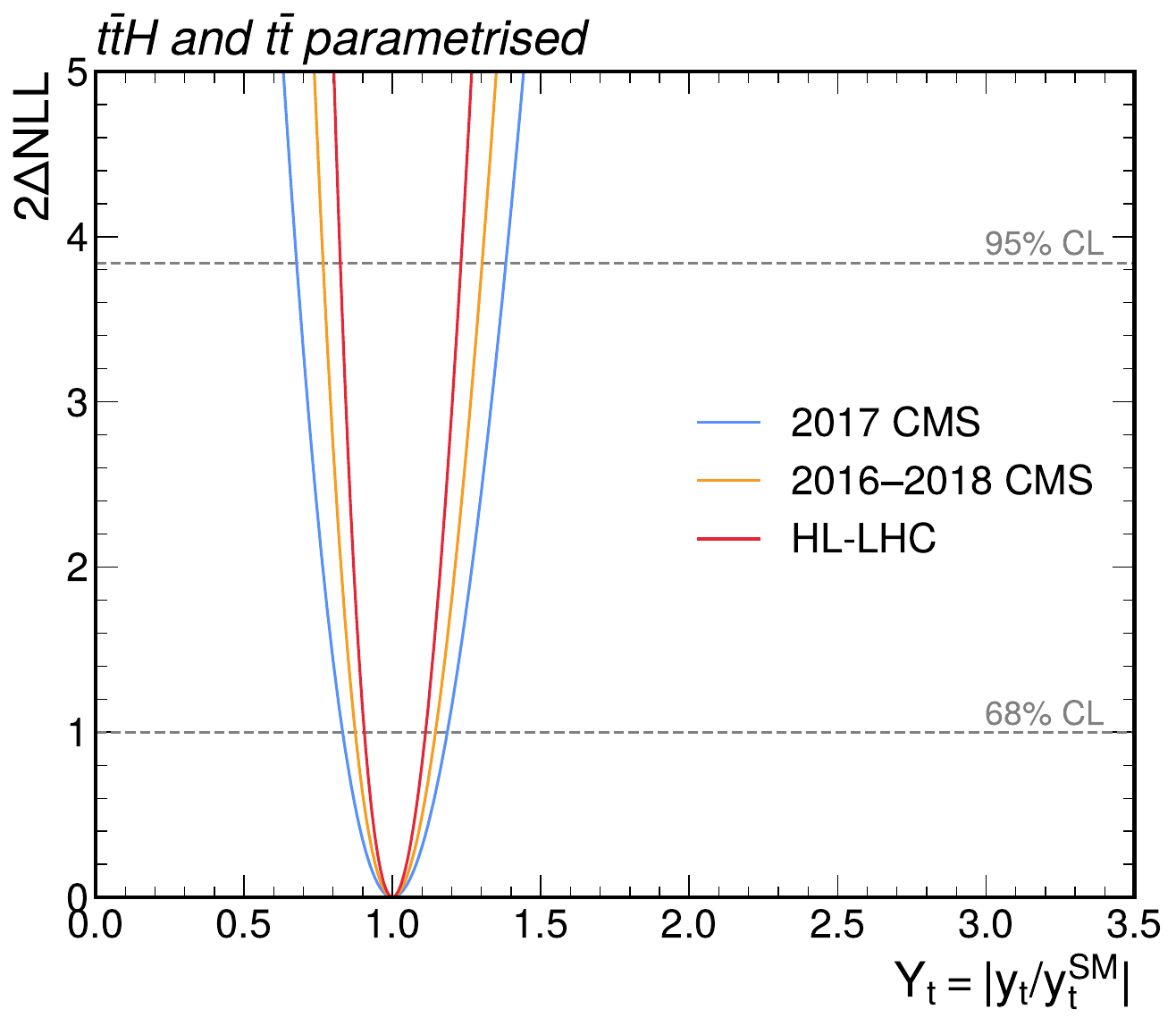}
        \caption{\ttH and \ttbar parametrized}
        \label{fig:nll-ttH-parametrised}
    \end{subfigure}
    \begin{subfigure}[b]{0.45\linewidth}
        \centering
        \includegraphics[width=\textwidth]{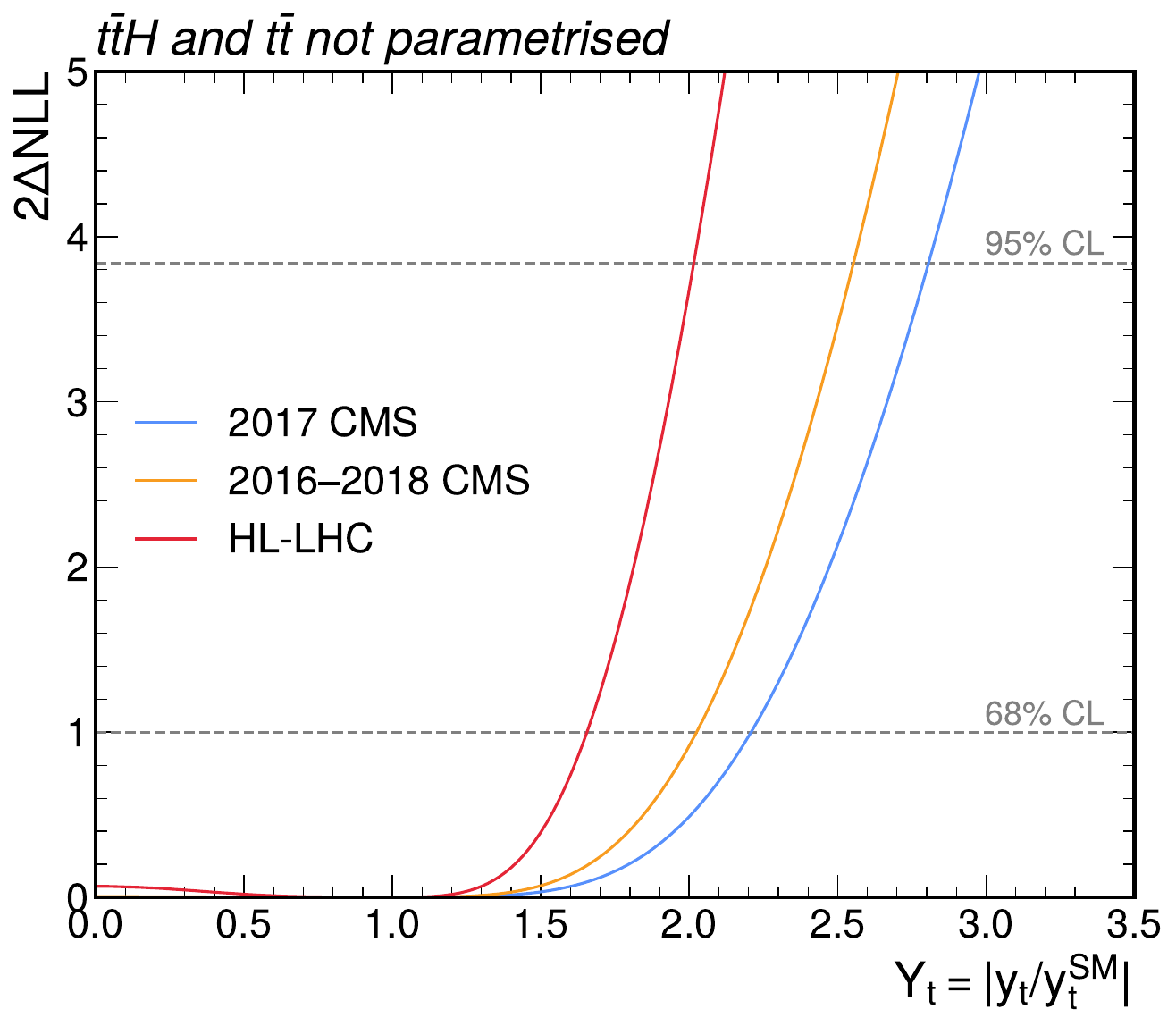}
        \caption{\ttH and \ttbar \emph{not} parametrized}
        \label{fig:nll-ttH-not-parametrised}
    \end{subfigure}
    \caption{Negative log-likelihood (NLL) scan of $Y_t$, showing the probable inferred range of the parameter up to 95\% confidence level. The parameter range derived at 68\%  and 95\% confidence levels is obtained from the region of the corresponding NLL value (shown in dashed lines) enclosed by the NLL scan curves. In Figure~\ref{fig:nll-ttH-not-parametrised}, the enclosed ranges should start at $Y_t = 0$ and end at the intersection between the NLL value and the scan curve.}
    \label{fig:nll-ttH}
\end{figure*}

When both \ttH and \ttbar processes are not parametrized to $Y_t$, and \ttH normalization is treated as a free parameter, the upper bounds inferred for the parameter are shown in Table~\ref{tab:yt-results-ttH-not-parametrised}. In this case, the lower bounds of $Y_t$ can not be inferred, corresponding to the negative log-likelihood scans shown in Figure~\ref{fig:nll-ttH-not-parametrised}. By observing the differences in negative log-likelihood scans for the case where both \ttH and \ttbar are parametrized to $Y_t$ (shown in Figure~\ref{fig:nll-ttH-parametrised}) and not parametrized, it is clear that the smaller range of the inferred values of $Y_t$ is heavily based on the parametrization of \ttH and $t\bar{t}$.

\begin{table}
    \centering
    \caption{Upper bounds of inferred $Y_t = |y_t/y_t^\textrm{SM}|$ at different data amounts, where the \ttH and \ttbar processes are not parametrized per $Y_t$}
    \label{tab:yt-results-ttH-not-parametrised}
    \begin{tabular}{c|cc}
        \hline
         Data amount    & 68\% CL & 95\% CL \\ \hline
         2017 CMS       & 2.208 & 2.807 \\
         2016--2018 CMS & 2.025 & 2.553 \\
         HL-LHC         & 1.655 & 2.015 \\ \hline
    \end{tabular}
\end{table}

\begin{figure}
    \centering
    \includegraphics[width=0.5\linewidth]{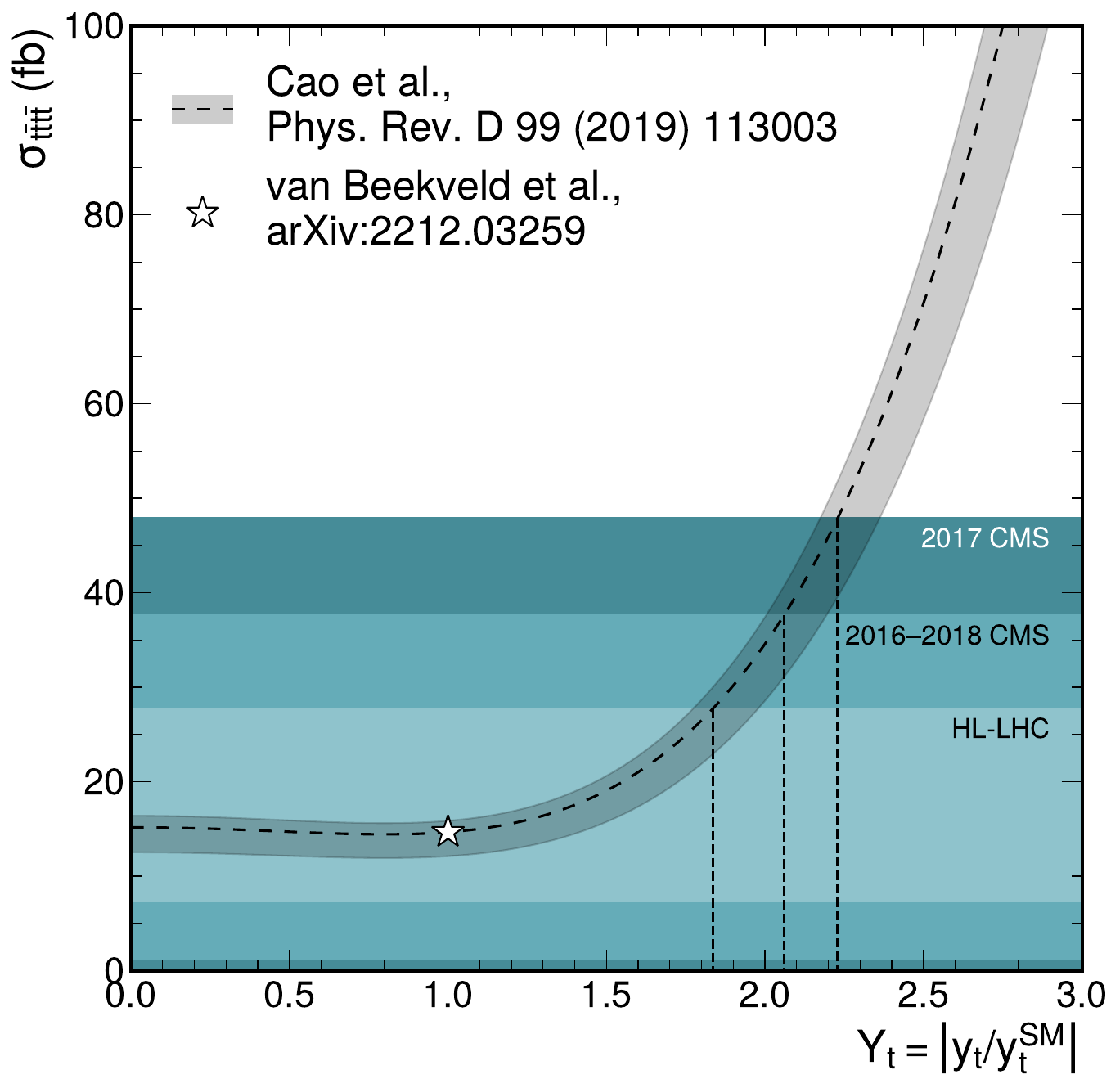}
    \caption{\fourtops cross section as a function of top Yukawa coupling ratio $Y_t = |y_t/y_t^\textrm{SM}|$,  based on the theoretical cross section prediction from Refs.~\cite{Cao_2019} and~\cite{vanbeekveld2025thresholdresummationproductionquarks}. Different color bands represent the inferred range of cross section at 2017 CMS, 2016--2018 CMS, and HL-LHC data amounts.}
    \label{fig:tttt-cross-section-translation}
\end{figure}

As explained in Section~\ref{subsec:template-histogram-xsect-measurement}, we can benchmark the above results with the $Y_t$ inference based on the inferred \fourtops cross section by comparing with the predictions provided in Equation~\ref{eqn:fourtops-yield-yt}, shown in Figure~\ref{fig:tttt-cross-section-translation}. The inferred cross-section values, along with the upper bounds of $Y_t$ obtained via the surrogate quantity, are shown in Table~\ref{tab:yt-results-cross-section}.

\begin{figure}
    \centering
    \includegraphics[width=0.5\linewidth]{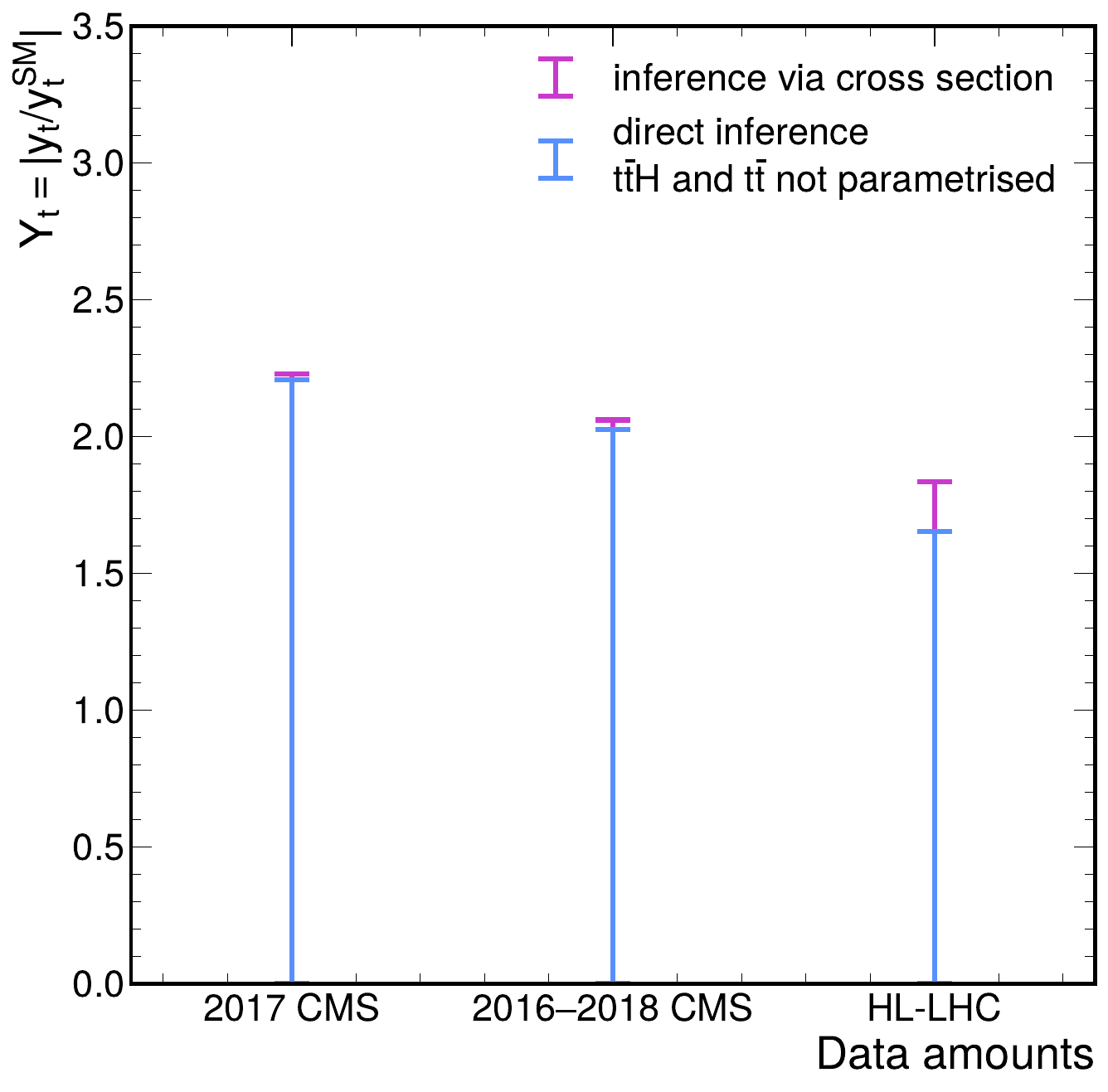}
    \caption{Comparison of inferred $Y_t$ range at 68\% confidence level between the direct inference and traditional inference via cross section, where \ttH and \ttbar normalization is not parametrized to $Y_t$}
    \label{fig:yt-ttH-vs-cross-section}
\end{figure}

The comparison of $Y_t$ inference between the traditional method and the direct inference method, where \ttH and \ttbar are not parametrized by $Y_t$, is shown in Figure~\ref{fig:yt-ttH-vs-cross-section}. As seen from the figure, the $Y_t$ upper bound at the 68\% confidence level is better when the parameter is directly inferred, even though the information from both \ttH and \ttbar processes is not involved.

%This improvement suggests that by applying neural network training that exploits kinematic changes for \fourtops process, the coupling measurement can be slightly improved compared to the method where the measured \fourtops cross section is compared.

This improvement over the inferred range of $Y_t$ suggests that our approach of exploiting extra simulation-level kinematic information for model learning can improve the inference of the parameter of interest in a physics model, compared to the traditional inference method that does not rely on them and is performed via a surrogate quantity. Furthermore, compared to the adaptation performed in Ref.~\cite{TOP-21-001}, the new inference approach relies on observation weights provided by the simulation instead of labeling datasets generated from different simulations as different classes. Another point to note is that the datasets used for the parameter inference in both methods have the same event selection requirement, which was originally designed for the discovery of \fourtops production and not specifically for the inference of $y_t$. Even though the event selection requirement is based on physics objects in each event, which is in turn generated from the simulation chain, we expect the qualitative conclusion to remain robust under reasonable variations of the event selection. 
%we do not expect this conclusion to change if the event selection requirements change.

\begin{table*}
    \centering
    \caption{Inferred \fourtops cross section and upper bounds of $Y_t$ at 68\% and 95\% confidence level}
    \label{tab:yt-results-cross-section}
    \begin{tabular}{lcccc}
        \hline
        & \multicolumn{2}{c}{68\% CL} & \multicolumn{2}{c}{95\% CL} \\
         Data amount    & cross section (fb) & $Y_t$ limit & cross section (fb) & $Y_t$ limit\\ \hline
         2017 CMS       & $14.65^{+33.23}_{-14.65}$ & 2.229 & $14.65^{+95.13}_{-14.65}$ & 2.819 \\
         2016--2018 CMS & $14.65^{+22.95}_{-13.40}$ & 2.060 & $14.65^{+65.50}_{-14.65}$ & 2.590 \\
         HL-LHC         & $14.65^{+13.05}_{-7.36}$  & 1.835 & $14.65^{+36.09}_{-10.96}$ & 2.270 \\ \hline
    \end{tabular}
\end{table*}

\section{Extension to multi-parameter inference: a CP-violation case study}
The summary statistics constructed in Section~\ref{subsec:template-histogram-construction} can also be used for interpretations of charge-parity (CP) symmetry violation. CP symmetry violation, CP-violation for short, is an extension to the SM postulating that the behavior of particles may be altered once both charge conjugation and parity operations are applied to the system. This type of symmetry has been shown to be broken in other particle productions before~\cite{PhysRevLett.13.138}, and measuring the CP-asymmetry in terms of top Yukawa coupling may give insights for certain theories beyond the Standard Model (BSM). 

In a purely computational sense, this postulate can modify the physics model in several ways, where the simplest modification is the addition of an extra parameter to the model. The interpretation for this postulate can then be performed by inferring the values of the extra parameters, in addition to the existing parameters in the model. For our case with $y_t$ and \fourtops production, it has been shown in a previous theoretical analysis~\cite{Cao_2019} that the \fourtops cross section can be altered by two parameters of interest, $a_t$ and $b_t$, representing CP-even and CP-odd top Yukawa couplings, respectively. If the CP-violation in the top Yukawa coupling does not exist, $b_t$ should be zero, and the coupling can be translated as $a_t = Y_t = |y_t/y_t^\textrm{SM}|$. The theoretical analysis performed in Ref.~\cite{Cao_2019} is similar to the inference via a surrogate quantity performed in Section~\ref{subsec:template-histogram-xsect-measurement}, where the upper bounds of the cross-section quantity are used to infer the upper bounds of both parameters $a_t$ and $b_t$ introduced by the extension.

In this work, however, we will use the summary statistics constructed for the inference of one parameter $y_t$ to infer the extended parameters $a_t$ and $b_t$. To achieve this, the summary statistics must be modified such that they are parametrized by both parameters. This will allow the inference over their continuous ranges, similar to the direct inference performed in Section~\ref{subsec:direct-yt-inference}. For simplicity, we will refer to the parameter $Y_t = |y_t / y_t^\textrm{SM}|$ as $a_t$ in this section.

The event yields for \fourtops and \ttH productions calculated from Section~\ref{subsec:template-histogram-construction} can be adjusted directly by the following factors~\cite{Cao_2019}:
\begin{eqnarray}
    %\mu_{t\bar{t}t\bar{t}}(a_t, b_t) &=& (7.724 - 1.164\,a_t^2 + 2.434\,b_t^2 + 0.910\,a_t^4 + 2.183 \, a_t^2 b_t^2 \nonumber \\
    %&& + 1.424\,b_t^4 ) / (7.724 - 1.164\,a_t^2 + 0.910\,a_t^4 ) \label{eqn:fourtops-yield-at-bt} \\
    \mu_{t\bar{t}t\bar{t}}(a_t, b_t) &=& \frac{7.724 - 1.164\,a_t^2 + 2.434\,b_t^2 + 0.910\,a_t^4 + 2.183 \, a_t^2 b_t^2 + 1.424\,b_t^4}{7.724 - 1.164\,a_t^2 + 0.910\,a_t^4} \label{eqn:fourtops-yield-at-bt} \\
    \mu_{t\bar{t}H}(a_t, b_t) &=& a_t^2 + 0.46 b_t^2 \label{eqn:tth-yield-at-bt}
\end{eqnarray}

Since the yields for \fourtops production in the template histogram are already adjusted in the SM behavior by $a_t$ (which is the same as $Y_t$), the denominator for the scaling factor above is calculated based on $a_t$ to convert to an arbitrary CP case where $b_t$ is non-zero. Furthermore, since both $\mu_{t\bar{t}t\bar{t}}$ and $\mu_{t\bar{t}H}$ are adjusted in terms of $a_t^2$, $b_t^2$, $a_t^4$, $b_t^4$, and $a_t^2b_t^2$, the yields for both processes should have the same effect whether both parameters are positive or negative. Hence, the inference for both parameters will be performed over the positive regions only. Furthermore, we will assume that \ttbar production is not affected by the extended parameter $b_t$.

By adjusting the event yields of both \fourtops and \ttH processes per $a_t$ and $b_t$, we can infer possible regions of both parameters within a certain confidence level. For this scenario, a likelihood-based inference over the possible regions of $a_t \in [0, 3.5]$ and $b_t \in [0, 3.5]$ is performed in a similar way to the direct inference of $y_t$.

\begin{figure*}
    \centering
    \begin{subfigure}[b]{0.45\linewidth}
         \centering
         \includegraphics[width=\textwidth]{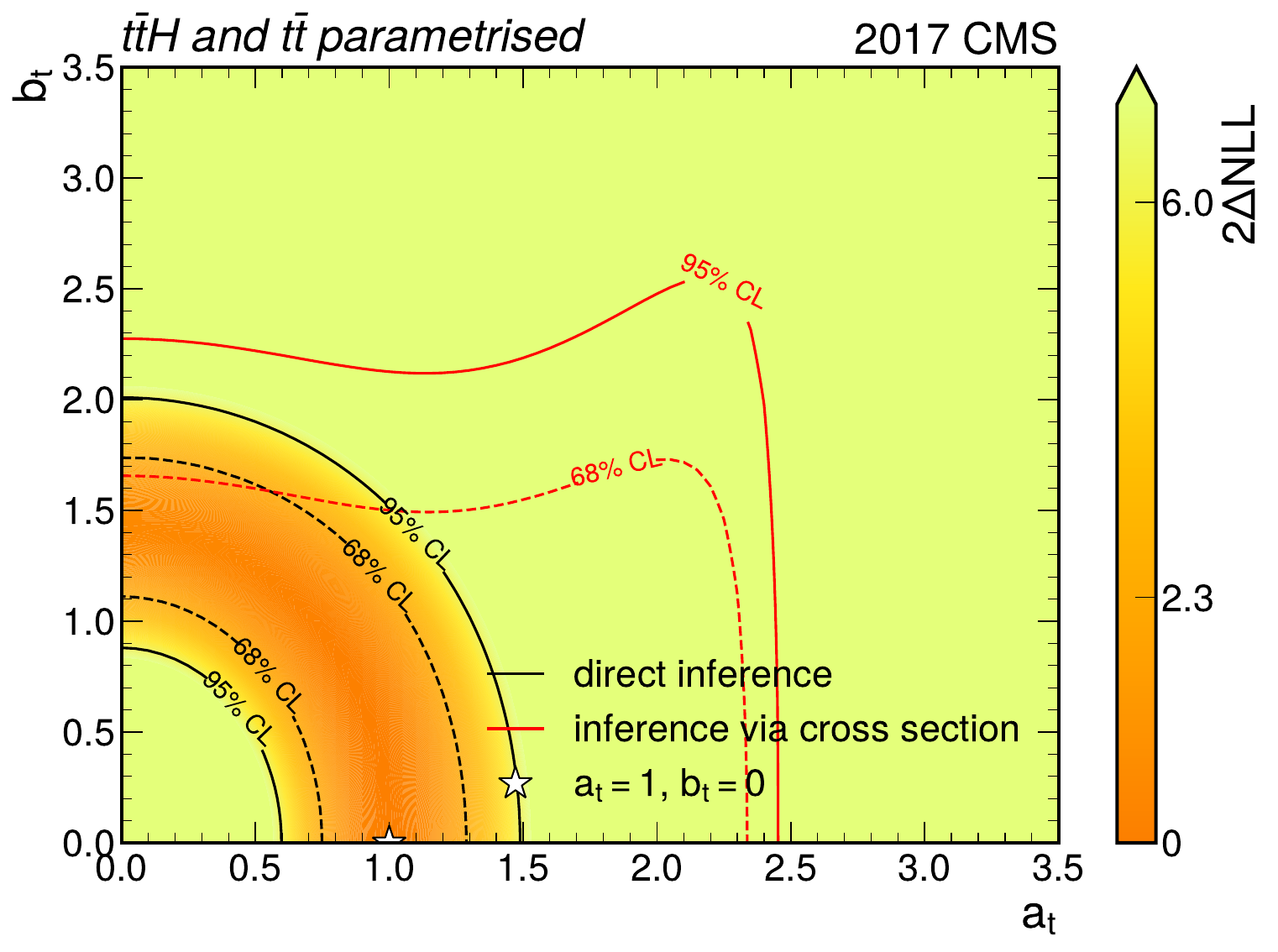}
         \label{fig:yt-vs-bt-2017}
    \end{subfigure}
    \begin{subfigure}[b]{0.45\linewidth}
         \centering
         \includegraphics[width=\textwidth]{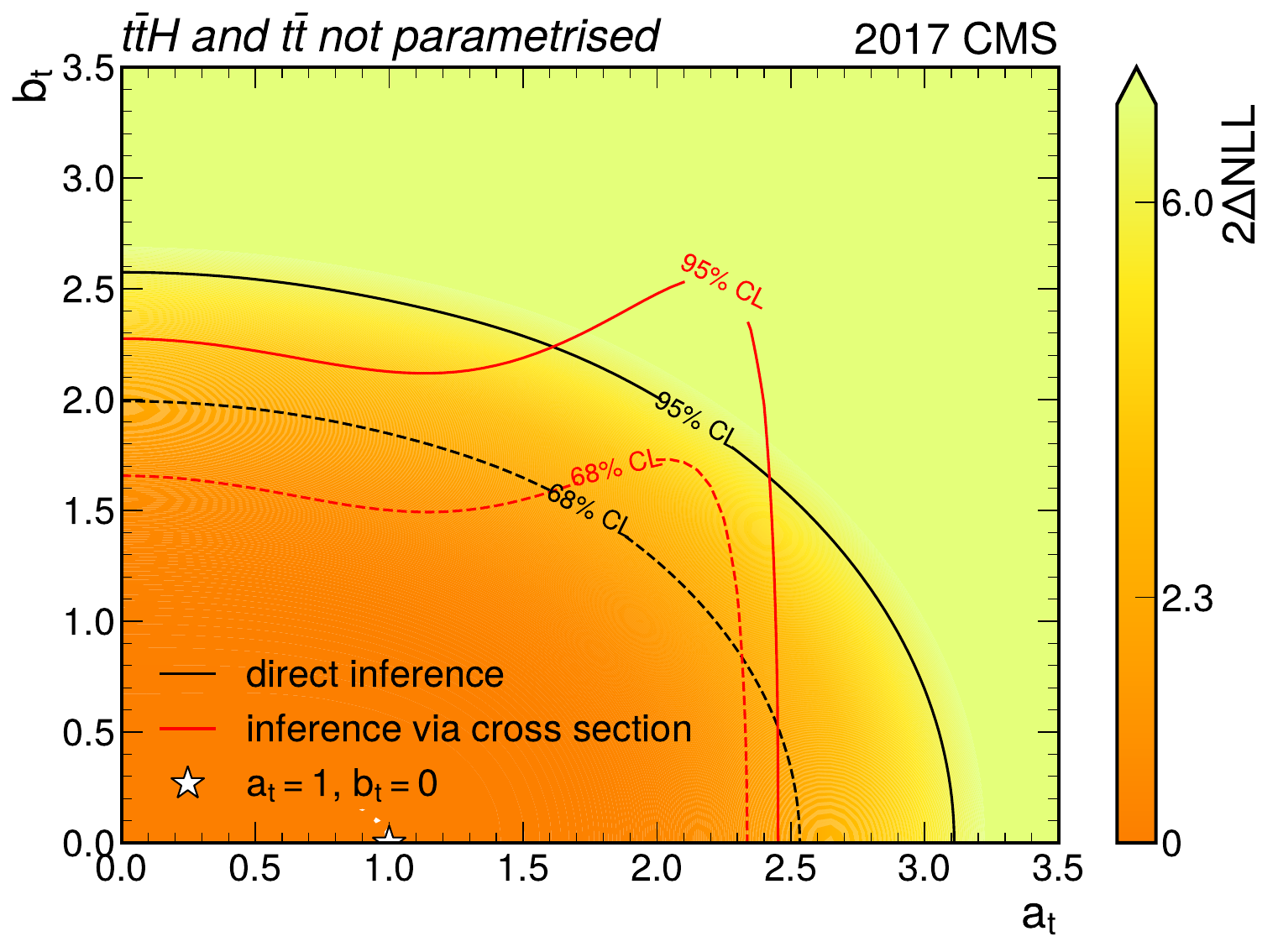}
         \label{fig:yt-vs-bt-2017-not-parametrised}
    \end{subfigure}
    \begin{subfigure}[b]{0.45\linewidth}
         \centering
         \includegraphics[width=\textwidth]{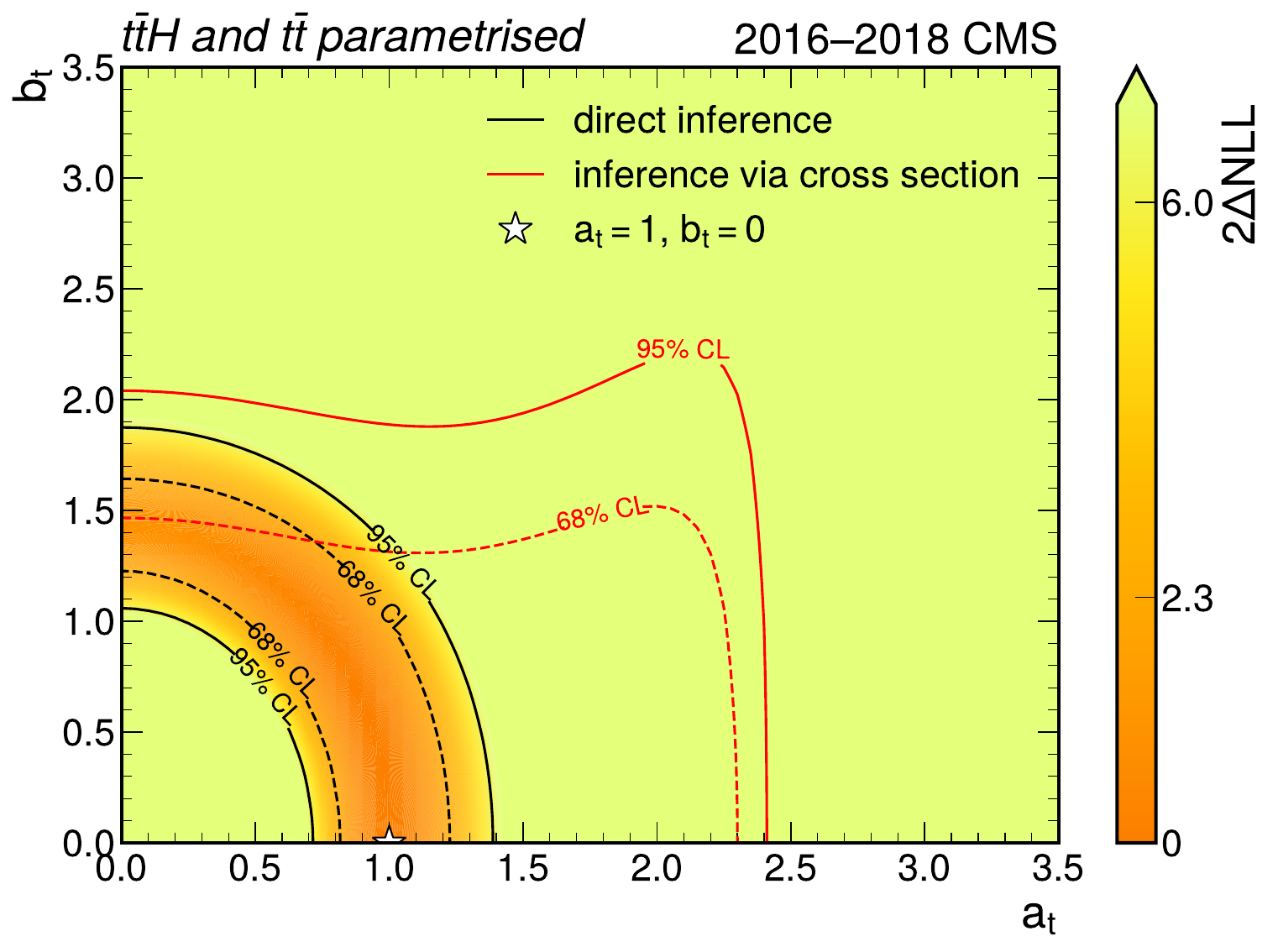}
         \label{fig:yt-vs-bt-run2}
    \end{subfigure}
    \begin{subfigure}[b]{0.45\linewidth}
         \centering
         \includegraphics[width=\textwidth]{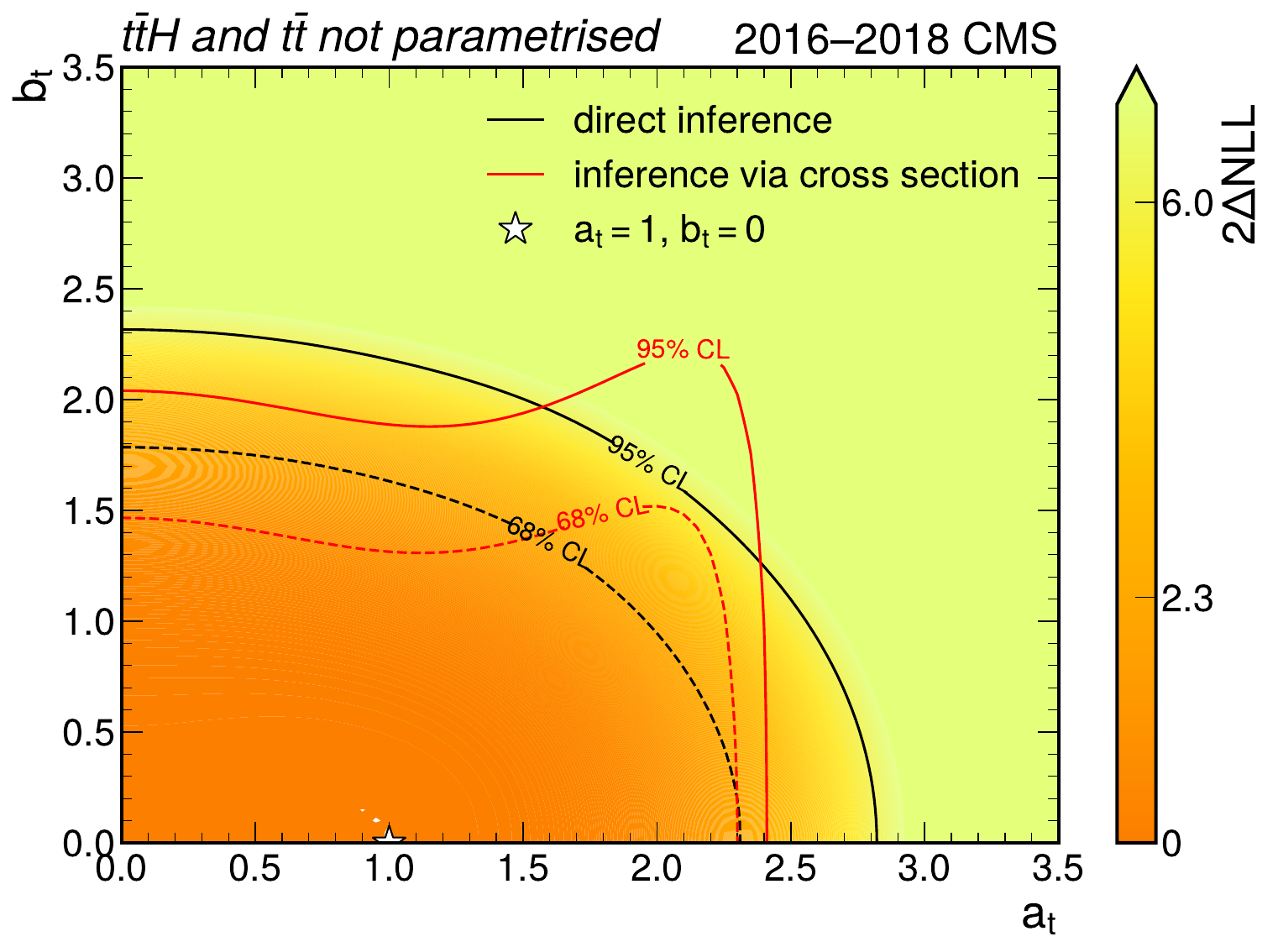}
         \label{fig:yt-vs-bt-run2-not-parametrised}
    \end{subfigure}
    \begin{subfigure}[b]{0.45\linewidth}
         \centering
         \includegraphics[width=\textwidth]{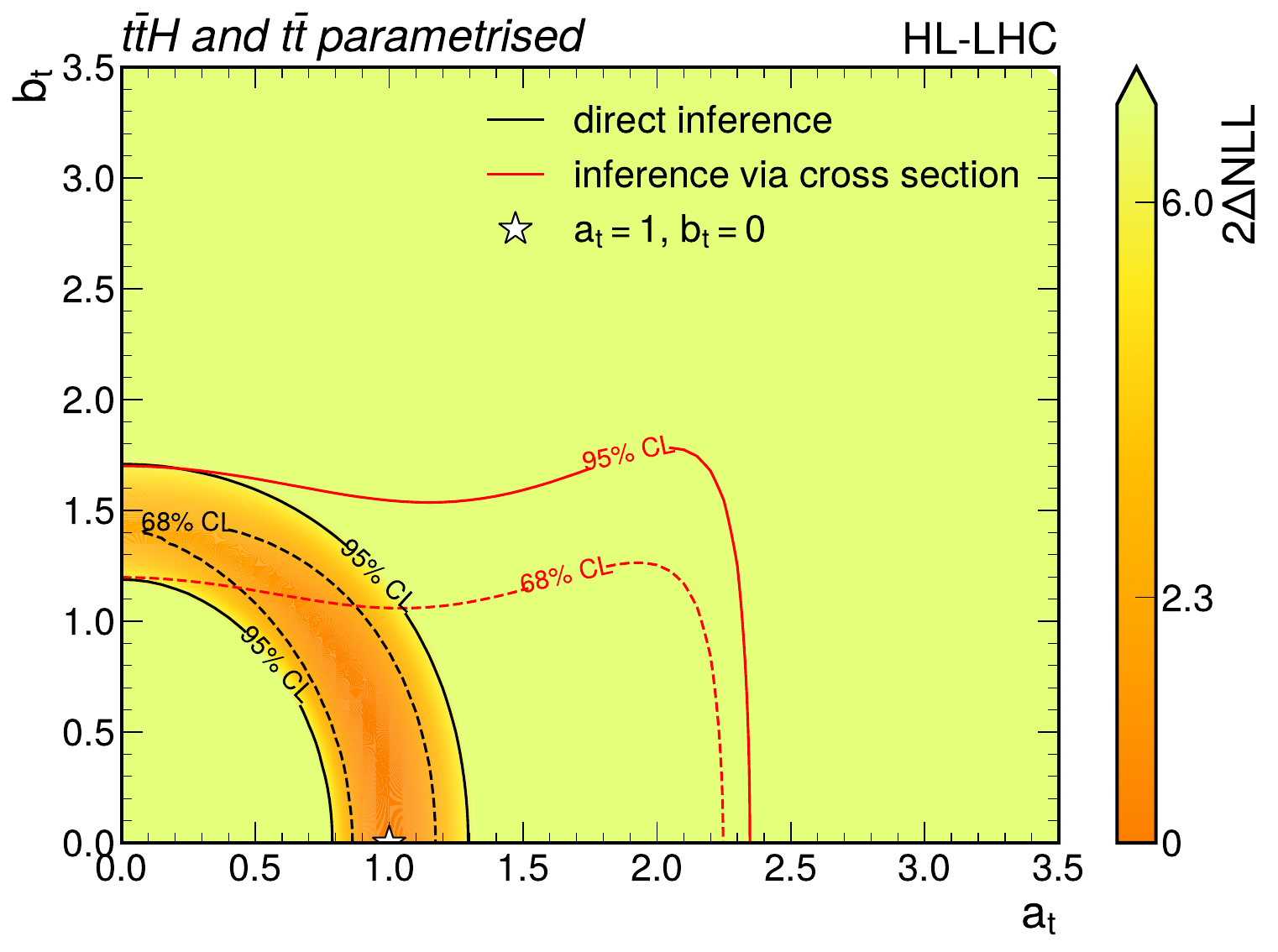}
         \label{fig:yt-vs-bt-3000}
    \end{subfigure}
    \begin{subfigure}[b]{0.45\linewidth}
         \centering
         \includegraphics[width=\textwidth]{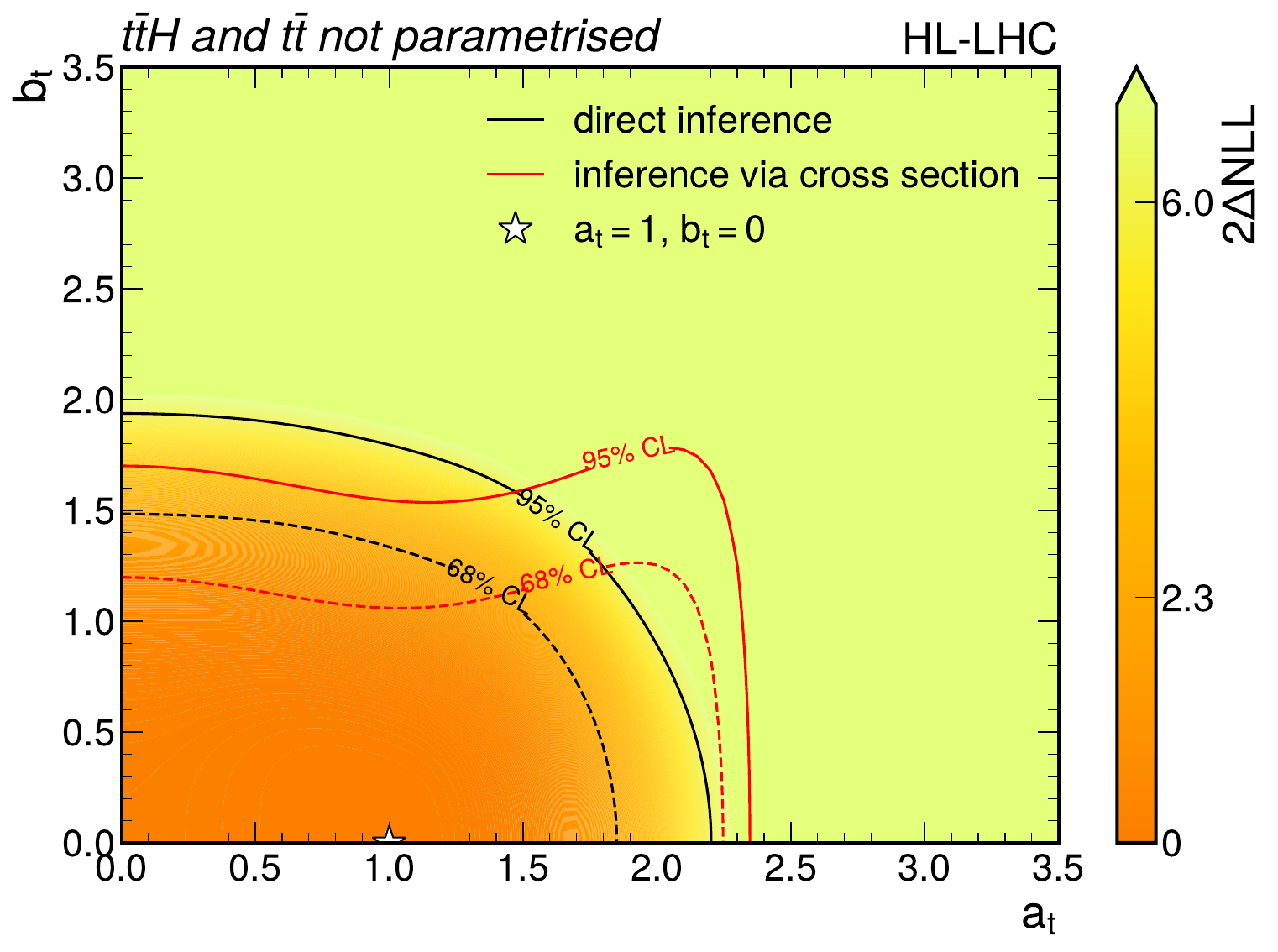}
         \label{fig:yt-vs-bt-3000-not-parametrised}
    \end{subfigure}
    \hfill
    \caption{Inferred regions $a_t$ and $b_t$ where \ttH and \ttbar processes are parametrized (left column) and \emph{not} parametrized (right column). Black lines represent regions confined by the parametrization of $Y_t$, while red lines represent regions confined by the measured cross-section values of \fourtops shown in Table~\ref{tab:yt-results-cross-section}. Dashed and solid lines represent enclosed regions at 68\% and 95\% confidence levels, respectively.}
    \label{fig:yt-vs-bt-ttH-parametrised}
\end{figure*}

The left column of Figure~\ref{fig:yt-vs-bt-ttH-parametrised} shows the inferred regions of the coupling values at the data amounts of 2017 CMS, 2016--2018 CMS, and HL-LHC, respectively. The expected parameter regions shown in the figure show that we can constrain the lower bounds of both parameters, both at 68\% and 95\% confidence levels.

The inferred \fourtops cross section, presented in Section~\ref{subsec:template-histogram-xsect-measurement}, can also be used as a surrogate quantity to infer the regions of both extended parameters, based on the altered yields in Equation~\ref{eqn:fourtops-yield-at-bt}. As shown in Figure~\ref{fig:yt-vs-bt-ttH-parametrised}, the inferred regions determined by the cross-section range are much larger than the inferred regions determined by the direct inference, where both \ttH and \ttbar processes are parametrized by both $a_t$ and $b_t$. When both processes are not parametrized, however, the inferred regions from the direct inference became larger than the one inferred via cross section, especially with lower amounts of data, as shown in the right column of Figure~\ref{fig:yt-vs-bt-ttH-parametrised}. At the data amount equivalent to HL-LHC, the inferred regions determined by the direct measurement become small enough to be comparable to the regions confined by the cross-section value alone. 

This improvement induced by our approach indicates that, by exploiting event weights from the simulation level, we can also apply the summary statistics constructed by this approach to studies of other domains where the inference models did not directly learn and still obtain comparable, if not improved, results compared to the inference via a surrogate quantity.

%It is crucial to note that, as this measurement is performed on two free parameters, $a_t$ and $b_t$, the regions confined may not translate directly to the one-parameter measurements presented in Section~\ref{sec:measured-results}.

\section{Summary and future prospects}
In this work, we have illustrated that, by incorporating weights calculated for observations by simulators into the inference model learning procedure, we can improve the inference of certain model parameters compared to the inference via a surrogate quantity alone. This is illustrated by the inference of top Yukawa coupling, the parameter of interest, from the simulation of \fourtops process, where our inference model learns the relation between input features and weights assigned to simulated observations. Observation weights from the simulator are calculated with respect to the parameter of interest, sampled over the continuous range, and allow our inference approach to infer the parameter value. The inferred range of the parameter by this approach shows an improvement over the inferred range by another traditional approach, where the parameter is inferred via a surrogate quantity.

Additionally, we can use our inference approach for the extension of the same model. Here, our inference model, which learns from the simulation based on one parameter of the model, is used in the inference for two parameters in the extended version of the model. Our results show that, provided enough amount of data, our inference approach can outperform the traditional inference approach, even though the inference model does not directly learn the relationship between input features and the extended parameters.

One thing to note in our approach is that we are utilizing the observable weight obtained from the simulator, which is easier to obtain compared to other inputs required by other approaches presented in Ref.~\cite{Brehmer_2018}. Weights attached to observables can be calculated within the simulator and are easy to obtain from the dataset simulated from the full simulator chain, mimicking particle collisions and detector simulations. On the other hand, some other inputs, as explained in Ref.~\cite{Brehmer_2018}, may require tinkering within the simulator chain in order to obtain such information only available from one part of the chain itself. Hence, our adapted approach can be easily applied to any parameter inference if the simulator has the capability of weight calculation right out of the box.

It should be obvious by now that this inference method should benefit from more simulations, either from the simulator itself or by expanding the regions of observations offered by the simulator. Furthermore, we believe that more sophisticated Machine Learning techniques tailored to utilize unique properties of input features from the observation will improve the inference capabilities. In the Particle Physics case, there are several models that leverage such unique properties, such as SPANet~\cite{10.21468/SciPostPhys.12.5.178}. Model-agnostic approaches based on unbinned likelihood, as suggested by Ref.~\cite{Brehmer_2018}, can also improve the exploitation of information available only from the simulation, such as observation weights, and in turn further improve parameter inference based on the simulated observations.

%\begin{figure}
% \centering
%        \includegraphics[width=0.5\textwidth]{figure1}
% \caption{Text describing the figure and the main conclusions drawn from it. To make your figures accessible to as many readers as possible, try to avoid using colour as the only means of conveying information. For example, in charts and graphs use different line styles and symbols. Further information is available in the online guide: \href{https://publishingsupport.iopscience.iop.org/publishing-support/authors/authoring-for-journals/writing-journal-article/\#figures}{https://publishingsupport.iopscience.iop.org/publishing-support/authors/authoring-for-journals/writing-journal-article/\#figures}}
%\label{fig1}
%\end{figure}

%\begin{table}
%\caption{Caption text describing the table. Adapt the template table below or replace with a new table. To add more tables, copy and paste the whole {\tt \textbackslash begin\{table\}...\textbackslash end\{table\}} block.}
%\centering
%\begin{tabular}{l c c c}
%\hline
%Column heading & Column heading & Column heading & Column heading \\
%\hline
%Data row 1 & 1.0 & 1.5 & 2.0 \\
%Data row 2 & 2.0 & 2.5 & 3.0 \\
%Data row 3 & 3.0 & 3.5 & 4.0 \\
%\hline
%\end{tabular}
%\label{tab1}
%\end{table}

%
% Each of the commands below will create an unnumbered section with the appropriate heading.
% Remove any sections that are not relevant for your article.
% All sections except suppdata will be removed if the [anonymous] option is used.
% See iopjournal-guidelines.pdf for more information.
%

\section*{Acknowledgment}
V. Wachirapusitanand was supported by the Program Management Unit for Human Resources \& Institutional Development, Research and Innovation, grant B13F680075. N. Srimanobhas was supported by the National Science Research and Innovation Fund program IND\_FF\_68\_369\_2300\_097. We acknowledge the supporting computing infrastructure provided by CU, CUAASC, and NSRF via PMUB B39G680009, and Thailand Science Research and Innovation Fund Chulalongkorn University ST\_69\_002\_2300\_001.

\bibliographystyle{ieeetr}  % or alpha, ieee, apalike, etc.
\bibliography{paper-ref}

\appendix
\section{Technical details on neural networks trained in this work} \label{app:nn-techincal-details}
In this appendix, the detailed structures for both the background rejection network and the parameter inference network are presented. The training for both networks is performed using Keras~\cite{keras}.

\subsection{Background rejection network} \label{subapp:bg-rejection-network-details}
The overall network structure contains the following neural network layers:
\begin{itemize}
    \item Input layer accepting a set of 56 input features
    \item Batch normalization layer
    \item A number of feed-forward dense layers (number of layers and neurons per layer adjusted during hyperparameter tuning), using ReLU activation function, each immediately followed by a dropout layer with a dropout probability of 0.1
    \item Output layer containing three nodes (representing $t\bar{t}t\bar{t}$, $t\bar{t}$, and \ttH processes), with softmax activation function
\end{itemize}

The training datasets for this network also include an additional dataset of \ttH production requiring the pair of top quarks to decay into two leptons and two neutrinos. This additional dataset is used during the training only and is not used in subsequent parameter inference steps.

Training datasets for this network contain 150\,000 events from each process to ensure equal representation during the training, and the remaining events for each process are put into testing datasets. The network is trained using categorical cross-entropy loss and the Adam optimizer with a learning rate of 0.05. To prevent overtraining, early stopping criteria are applied where the training will stop if there are no improvements over the validation loss for five consecutive epochs, and the model with the best validation loss is saved during the training.

Hyperparameter tuning is used to determine the optimal number of layers and the number of neurons per layer. The goals of this tuning are to give the highest amount of \fourtops signal events versus both background events when the cut of 0.6 at the \fourtops output node is applied, while at the same time providing the continuous output distribution spanning in the range of $[0, 1]$.

To ensure that the output of all nodes is continuous, events from both training and testing datasets are populated based on the output of \fourtops and \ttH nodes. Since the sum of all the outputs in each event must be 1 due to the softmax activation function in the output layer, all events are separated into 55 categories, depicted in Figure~\ref{fig:classification-network-category}. Each category spans the output range (from \fourtops and \ttH nodes) of 0.1. For instance, the $t\bar{t}t\bar{t}$-rich region, shown at the top left part of the figure, must have the \fourtops output between $[0.9, 1.0)$ and \ttH output of $[0.0, 0.1)$. During hyperparameter training, a network configuration will be vetoed if there is at least one out of 55 event categories where no events are populated. The same 55 event categories will also be used to construct the template histograms for $y_t$ inference later.

The best configuration for this network, determined via hyperparameter tuning, contains three dense layers of 50, 100, and 100 neurons and a dropout probability of 0.1.

\subsection{Parameter inference network} \label{subapp:param-inference-network-details}
The overall network structure is similar to the background rejection network, as follows:
\begin{itemize}
    \item Input layer accepting 44 input features
    \item Batch normalisation layer
    \item A number of feed-forward dense layers (number of layers and neurons per layer adjusted during hyperparameter tuning), using $\tanh$ activation function, each immediately followed by a dropout layer with dropout probability to be tuned via hyperparameter tuning
    \item Output layer with one node, using sigmoid activation function
\end{itemize}

The training dataset for this network contains 250\,000 \fourtops events from each of the two event categories assigned by the weight ratio between two points of $y_t$ values, while the remaining events are gathered as the testing dataset. This network is trained using binary cross-entropy loss and Adam optimizer with a learning rate of 0.05. The same early stopping criteria is also applied, where the training will stop after three epochs of no improvement over the validation loss.

During the hyperparameter tuning for this network, a binning scheme for one particular configuration is determined such that each bin in the output distribution would have the highest variance as they are altered by $y_t$. The binning scheme is determined by separating the output distribution calculated from the training dataset into two halves from the point where the normalized count difference between low-weight events and high-weight events is lowest. Then, each half of the distribution is divided into three smaller sections with equal yields of the training events of low-weight type. This results in six histogram bins (as shown in Figure~\ref{fig:output-distribution-yt-optimal-cuts}), and the goal of hyperparameter tuning over this network is to determine the network configuration with the highest minimum change ratio (out of six histogram bins) compared between two points of $y_t$ values. The highest minimum change ratio guarantees that the optimal network will provide significant differences for event yields regardless of the histogram bins.

The optimal network from hyperparameter tuning contains four layers of 200, 200, 50, and 50 neurons, and a dropout probability of 0.2. Figure~\ref{fig:output-dist-train-test} illustrates the training output distribution from the optimal $y_t$ network, as determined by the hyperparameter tuning requirement. The figure also shows no signs of overtraining as the output distributions between the training and testing datasets are similar.

\end{document}